\documentclass[%
 reprint,
nofootinbib,
 amsmath,amssymb,
 aps,
prd,
floatfix,
]{revtex4-2}
\usepackage{xcolor}
\usepackage{graphicx}
\usepackage{dcolumn}
\usepackage{bm}
\usepackage{hyperref}

\usepackage[normalem]{ulem}

\graphicspath{{./Figures/}}

\usepackage{physics}

\usepackage{placeins}

\begin{document}

\preprint{APS/123-QED}

\title{Evolution of a black hole cluster in full general relativity}
\author{Jamie Bamber}
\affiliation{%
Department of Physics, University of Illinois Urbana-Champaign, Urbana, IL 61801, USA
}
\author{Stuart L. Shapiro}
\affiliation{%
Departments of Physics and Astronomy, University of Illinois Urbana-Champaign, Urbana, IL 61801, USA
}
\affiliation{National Center for Supercomputing Applications, University of Illinois Urbana-Champaign, Urbana, IL 61801, USA}
\author{Milton Ruiz}%
\affiliation{%
 Departament d’Astronomia i Astrof\'{i}sica, Universitat de Val\`{e}ncia, C/ Dr Moliner 50, 46100, Burjassot (Val\`{e}ncia), Spain
}
\author{Antonios Tsokaros}
\affiliation{%
Department of Physics, University of Illinois Urbana-Champaign, Urbana, IL 61801, USA
}%
\affiliation{National Center for Supercomputing Applications, University of Illinois Urbana-Champaign, Urbana, IL 61801, USA}
\affiliation{Research Center for Astronomy and Applied Mathematics, Academy of Athens, Athens 11527, Greece}

\date{\today}

\begin{abstract}
We evolve for the first time in full general relativity a small, {\it collisional} $N$-body black hole (BH) cluster of  arbitrary total mass $M$.
The bound cluster is initially compact (radius $R/M \sim 10$), stable, and consists of  
25 equal-mass, nonspinning BHs. The dynamical interactions of compact objects in $N$-body clusters is of great interest for the formation of BHs in the upper mass gap as well as intermediate and supermassive BHs. These are potential sources of gravitational waves that may be detected by both current and future observatories. Unlike previous $N$-body Newtonian and post-Newtonian simulations, no ``subgrid physics" is required to handle collisions and mergers. We can therefore confirm in full general relativity several predictions from these simulations and analytic estimates: the runaway growth of a large BH via repeated mergers; spindown of the central BH with increasing captures; the ejection of a BH with a large asymptotic velocity due to a several-body interaction; and a regime where mergers occur primarily via direct collisions on highly eccentric orbits instead of quasicircular inspirals. We extract the gravitational wave signal and find it has several distinct features associated with the compact cluster regime. Our results suggest the signal is sufficiently loud that next generation observatories would likely be able to detect similar events across most of the observable universe. This work is a preliminary proof-of-principle study that we hope will open up a new arena for numerical relativity and the study of $N$-body compact systems.

\end{abstract}

\maketitle


\section{Introduction}

\label{sec:astro_motiv}

The study of BHs in compact clusters is of interest to a broad range of astrophysical issues. First there is the question of the origin of the supermassive BHs (SMBHs) found at the centers of galaxies. These have masses $10^6 - 10^{10} M_{\odot}$, compared to the typical mass of $\sim 5-50 M_{\odot} $ for stellar-origin BHs formed from the collapse of massive stars \cite{Remillard:2006fc,KAGRA:2021vkt}. Observations of quasars at redshifts $> 7$ \cite{Banados:2017unc,Eilers:2024xus,Greene:2024phl} indicate that SMBHs must have formed in the first billion years of the universe's existence, making the origin of these massive BHs a key astrophysical puzzle. One possible scenario is growth from intermediate mass BH (IMBH) seeds of mass $\sim 10^{2} - 10^{5} M_{\odot}$ \cite{Volonteri:2021}. There is little direct evidence of IMBHs to date \cite{Greene:2020} compared to the other BH populations. The heaviest BH formed from a binary BH merger observed by LIGO, the $\sim 142 M_{\odot}$ daughter BH in GW19052 \cite{LIGOScientific:2020ufj}, and the lightest BH candidate at the center of a galaxy, the $\sim 6800 M_{\odot}$ object in the dwarf elliptical galaxy NGC 205 \cite{Nguyen:2019}, both fall within the IMBH mass range. Recent observations of fast moving stars in $\omega$Centauri, the Milky Way's most massive globular cluster, have also indicated the presence of a IMBH with a mass greater than 8200$ M_{\odot}$ \cite{Haberle:2024}.

IMBHs are important not only for galaxy formation channels in the early universe but also as sources of gravitational waves observable by LISA (see e.g. \cite{Miller:2008fi,Strokov:2023kmo}) and other next-generation detectors. One possible formation channel for IMBHs \cite{Giersz:2015,Greene:2020,Volonteri:2021} involves gravitational runaway in dense stellar environments \cite{Mapelli:2021syv,Mapelli:2021gyv}, such as nuclear star clusters (NSCs) at the centers of galaxies \cite{Antonini:2016gqe, Antonini:2018auk,Fragione:2020nib,Sedda:2020jvg,Chattopadhyay:2023pil,Gaete:2024ovu} and massive globular clusters (GCs) \cite{PortegiesZwart:2000zi,Miller:2001ez,Tanikawa:2013,Rodriguez:2016kxx,Hong:2018bqs,Askar:2018,Choksi:2018jnq,Rizzuto:2021,Kamlah:2022,Fragione:2023kqv,Torniamenti:2024uxl}. As the masses of BHs formed (typically $\sim 5 - 27 M_{\odot}$ \cite{Belczynski:2005ai} for young stellar clusters) are larger than the average mass of a main sequence star, mass segregation via dynamical friction \cite{Chandrasekhar:1943,Spitzer:book} causes the BHs to sink to the center and form a compact subcluster (see e.g. \cite{Breen:2013vla} and Sec. \ref{sec:dyn_context} below). The detection of several BH candidates in globular cluster NCG 3201 \cite{Giesers:2018,Giesers:2019} and the presence of a large well-resolved core has been taken, via comparison with numerical simulations, to provide observational evidence of the presence of a $\gtrsim 200$ BH subsystem within the cluster \cite{Kremer:2018,Askar:2018}.

These BHs form binaries, which arise and harden via 3-body interactions and gravitational radiation \cite{Quinlan:1990,Heggie:2009}. Both 3-body super-elastic interactions (e.g. \cite{Banerjee:2009hs,Downing:2010,Kremer:2019rid}) and the anisotropic emission of gravitational waves during mergers (e.g. \cite{Peres:1962,Lousto:2009mf}) produce kicks which can terminate a hierarchical merger chain by ejecting daughter BHs and/or binaries from the cluster, depending on the cluster escape velocity \cite{Sigurdsson:1993zrm}. More massive BHs have larger inertia and are more likely to be retained: if such a heavy BH forms, either from the collapse of a massive star \cite{Miller:2001ez} or hierarchical mergers \cite{Rodriguez:2019huv,Gerosa:2021mno}, and the core is sufficiently dense \cite{Giersz:2015} it may undergo runaway growth via repeated mergers with lower mass BHs until it reaches IMBH mass scales of $\sim 10^3 M_{\odot}$ \cite{Miller:2001ez,Mouri:2002mc}. A similar process may occur with massive main-sequence stars \cite{Quinlan:1990,PortegiesZwart:2002iks}. 

The interactions of BHs and neutron stars (NSs) in dense cluster environments may thus be important as one of the possible formation channels for the population of $\sim 100$ binary merger GW events detected to date by LIGO-Virgo-KAGRA \cite{PortegiesZwart:1999nm,Downing:2010,Tanikawa:2013,Askar:2016jwt,Gerosa:2017kvu,Rodriguez:2019huv,Fragione:2020han,Gerosa:2021mno} and sources that will be detectable by LISA \cite{Kremer:2018cir,Arca-Sedda:2020lso}. In particular, hierarchical mergers of second (or higher) generation BHs \cite{Kimball:2020qyd,Mahapatra:2022ngs,Mahapatra:2024qsy}, formed from previous mergers, could account for binary components in the ``upper mass gap" from $\sim 40 - 120 M_{\odot}$ where stellar-origin first generation BHs are unlikely to occur due to pair-instability supernovae \cite{Gerosa:2021mno,Fragione:2023kqv,Torniamenti:2024uxl}. 
One scenario which could lead to a relativistic cluster of BHs is mass segregation followed by gravothermal core collapse on relaxation timescales \cite{Shapiro:1985c,Quinlan:1987,Quinlan:1990}. An additional formation scenario is hydrodynamical friction and accretion driving gravitational capture in a matter environment (such as AGN accretion disks \cite{McKernan:2012rf,Bartos:2017,McKernan:2020lgr,Wang:2024}). Under some models primordial BHs (PBHs), if they exist, may also form bound clusters \cite{Belotsky:2018wph,Trashorras:2020mwn,Stasenko:2021vmm,Franciolini:2022ewd,Siles:2024yym} with implications for the constraints on the upper bound of the primordial BH dark matter mass fraction \cite{DeLuca:2020jug}. PBHs may also be captured by NSs (see e.g. \cite{Capela:2013yf,Richards:2021upu,Baumgarte:2024iby,Baumgarte:2024ouj,Caiozzo:2024flz}), where the PBHs lose energy due to the dissipative interaction with the NS matter and are driven to the center. This could result in a bound relativistic PBH cluster within the star, provided the timescale for each PBH capture is much less than the timescale for dynamical collapse of the host NS. On the other end of the mass scale, the hierarchical merger of (proto)galaxies and their associated IMBHs and SMBHs can help to produce the massive SMBHs observed today (see e.g. \cite{Sobolenko:2021orc,Kritos:2024sgd}). The gravitational wave signals from such  mergers will be also be a key target for LISA \cite{New:2000dv,Hughes:2001ya,LISA:2017,LISA:2022kgy} and pulsar timing array (PTA) searches (e.g. \cite{Kelley:2017lek}). 

Dense clusters of compact objects will not only produce gravitational waves through mergers but also via gravitational bremsstrahlung in close hyperbolic and parabolic ``fly-by" encounters (see e.g. \cite{Kocsis:2006hq,Garcia-Bellido:2021jlq,Morras:2021atg,Kerachian:2023gsa}) which may be detectable with matched filtering \cite{Kocsis:2006hq,Morras:2021atg}, burst searches, or from their contribution to the stochastic gravitational wave background \cite{Garcia-Bellido:2021jlq,Kerachian:2023gsa}. 

In a new series of investigations we aim to address some key unanswered questions regarding these clusters by evolving, for the first time, a compact $N$-body cluster of BHs in full numerical GR. This project is a major extension of our earlier computational treatment of compact {\it collisionless} clusters, labelled as ``Relativistic Stellar Dynamics on the Computer" \cite{Shapiro:1985a,Shapiro:1985b,Shapiro:1985c,Shapiro:1986}, as we now move into the {\it collisional} domain and focus on BHs. As it is our initial foray, the system studied here is highly idealized and the reported results are very preliminary. Our real goal is to pave the way for more advanced and detailed studies to follow.

Unlike Newtonian and post-Newtonian simulations our $N$-body GR treatment requires no ``subgrid physics", nor
does it need to impose any ``regularization" when BHs orbit close to each other and/or merge. However, due to the long timescales involved, coupled with our finite computational resources, the systems tackled here are chosen to be highly compact ($R/M \sim 10$) bound clusters, with the BHs selected initially from a distribution function that yields a stable equilibrium system in the infinite particle, (i.e. smooth distribution) limit. Our resulting clusters contain a small number ($N=25$) of BHs that all begin with equal mass and no spin.

In Section II we detail the different dynamical regimes of self-gravitating $N$-body systems, results from past simulations, and the key questions we aim to explore. In Section III we estimate some characteristic timescales that govern the evolution of the cluster and exhibit their scaling with the adopted parameters. In Section IV we describe our numerical methods and simulation setup. In Section V we summarize our results and in Section VI our Conclusions. Throughout the paper we adopt geometrized units with $G=1=c$, unless otherwise noted. Times and lengths are given in terms of the cluster Arnowitt–Deser–Misner (ADM) mass $M = 5(M/10^6M_{\odot}) s = 1.5\times 10^{3}(M/10^6M_{\odot}) \mathrm{km}$.

\section{Dynamical regimes}
\label{sec:dyn_context}

The dynamics of a self-gravitating $N$-body system of particles (e.g., stars, BHs, dark matter, etc.) depends on some important timescales, which serve to distinguish the different physical regimes of the system \cite{Lightman:1978zz,Dehnen:2011fj}. These depend critically on several of the global parameters defining the system (e.g. the typical velocity dispersion $v$, the radius $R$,  mass $M$ and the number $N$ of constituents) and two of them (e.g. the velocity and radius) can always be related by the virial theorem for systems in dynamical equilibrium. The timescale estimates assembled below are largely based on Newtonian dynamics. 

The shortest timescale in a typical $N$-body cluster is the dynamical, or crossing, time,
\begin{equation} 
\label{tdyn}
 t_{\textup{dyn}} \sim \frac{R_h}{v},
\end{equation}
where $R_h$ is the radius containing half the mass. The next most important timescale is the 2-body relaxation time due to local fluctuations in the gravitational field. It is roughly the time it takes for the cumulative effect of many distant encounters (gravitational scatterings) to deflect a particle by $\sim 90^{\circ}$,
\begin{equation}  
  t_{\textup{rh}} \sim \frac{v^3}{15.4 m^2 n \ln(0.4N)},
\end{equation}
(see \cite{Spitzer:1971} Eqn. (5)) where $m$ is the particle mass and $n$ the local particle density\footnote{for a relativistic generalization of the relaxation timescale, see Shapiro et al. (2018) \cite{Shapiro:2018vju} Eqn. 45.}. This timescale measures the energy equilibration time for a large-$N$ system in dynamical equilibrium, i.e., the time it would take a homogeneous system to achieve a Maxwellian velocity distribution. The key ratio is 
\begin{equation}
    \frac{t_{\textup{dyn}}}{t_{\textup{rh}}} \sim \frac{26 \log_{10} (0.4 N)}{N},
\end{equation}
(see \cite{Spitzer:1971} Eqn. (6), \cite{Lightman:1978zz}).

\subsection{Collisionless clusters}

For $N \gg 1$, the above ratio is very small and large-$N$ clusters can be approximated as ``collisionless" to a high degree if their lifetimes are less than $t_{\textup{rh}}$. These clusters closely satisfy the collisionless Boltzman (Vlasoff) equation which is given by
\begin{equation}
    \pdv{f}{t} + \left(\dv{x^j}{t}\right)\pdv{f}{x^j} + \left(\dv{p_j}{t}\right)\pdv{f}{p_j} = 0,
    \label{eq:Vlasoff}
\end{equation}
where $f(t;x^j,p_j)$ is the particle distribution function in 6-dimensional phase space and ($x^j, p_j$) are the spatial components of the particle position and momentum, respectively (for a textbook treatment in general relativity, see, e.g., \cite{Baumgarte:2010}). Thus, for systems where the total lifetime is less than $t_{\textup{rh}}$ collisional relaxation is never important. This is the case for stars in most galaxies, for galaxies in clusters, and for truly collisionless dark matter particles (e.g. axions). The regime applies for smooth clusters of total rest mass $M_0 > 0$ in the limit where the particle mass obeys $m = M_0/N \rightarrow 0$ while $N \rightarrow \infty$. The numerical evolution of clusters in this collisionless limit was first probed in full general relativity by Shapiro \& Teukolsky in their series of papers ``Relativistic Stellar Dynamics on the Computer", for
which they adopted a mean-field, particle simulation scheme for both spherical \cite{Shapiro:1985a,Shapiro:1985b,Shapiro:1986} and axisymmetric systems \cite{Shapiro:1993b,Shapiro:1992b,Abrahams:1994ge,Hughes:1994ea,Shapiro:1995rr}. Their simulations tested binding energy criteria for the stability of equilibrium clusters, tracked the collapse to BHs of unstable systems, probed relativistic violent relaxation, explored the head-on collision of two BHs, each of which formed from cluster collapse, and, for nonspherical systems, examined the topology of merged event horizons, tested the hoop conjecture and determined the generation of GWs. These simulations also proved useful to help develop early on some of the tools of numerical relativity.

\subsection{Weakly collisional clusters}
\label{sec:weakly_coll}

For smaller $N$ the ratio $t_{\textup{dyn}}/t_{\textup{rh}}$ increases. If the lifetime of the cluster $t_{\textup{life}}$ is such that $t_{\textup{rh}} < t_{\textup{life}}$ then the evolution is influenced by collisional relaxation and is described by the collisional Boltzmann equation 
\begin{equation}
    \pdv{f}{t} + \left(\dv{x^j}{t}\right)\pdv{f}{x^j} + \left(\dv{p_j}{t}\right)\pdv{f}{p_j} = \Gamma[f],
\end{equation}
where $\Gamma$ is a functional determined by the encounter (gravitational scattering) rate. For sufficiently large $N$ the functional can be approximated by the Fokker-Planck approximation. Such systems can be termed ``weakly collisional". Relaxation via the cumulative effect of many distant encounters between particles drives the system towards relaxation and energy equipartition. This causes mass segregation where heavier particles sink towards the core and lighter particles are pushed to the outskirts, resulting in an energy flow from the contracting core to the extended halo \cite{Spitzer:1969,Spitzer:1971,Lightman:1978zz,Binney:2008,Dehnen:2011fj}. This process eventually leads to a ``gravothermal catastrophe" or ``secular core collapse" \cite{Anatov:1962,Lynden-Bell:1968,Henon:1971}: dramatic contraction of the cluster's innermost regions on relaxation timescales as the core evolves to higher and higher density and velocity dispersion, but with smaller and smaller core radius and particle number. This behavior applies to the evolution of stars in globular clusters and dense galactic nuclei, and potentially to halos of self-interacting dark matter (SIDM) \cite{Balberg:2002ue,Balberg:2001qg}. Gravothermal catastrophe of compact star clusters was first proposed as a possible formation mechanism for SMBHs in \cite{Shapiro:1985c} (see also \cite{Quinlan:1987,Quinlan:1990}). It was 
subsequently suggested that SIDM halos undergoing gravothermal collapse might also lead to SMBH and IMBH formation in \cite{Balberg:2002ue}, an idea pursued further in \cite{Pollack:2015,Shapiro:2018vju,Meshveliani:2022rih}. 

In addition to the relaxation timescale set by long-range 2-body interactions, short-range interactions can additionally form bound binaries. ``Hard" binaries are those whose binding energy exceeds the typical kinetic energy of surrounding particles, and, in general, further interactions with other particles cause hard binaries to become harder and soft binaries to become softer \cite{Heggie:1975}. For small-$N$ clusters ($10^2 - 10^3$ particles) these short-range close encounters become more important, the ratio of dynamical to relaxation time becomes larger, as does the energy of escaping stars \cite{Spitzer:1971}. The formation of binaries occurs more prominently, both by 3-body and, in the case of stars, dissipative tidal encounters, and direct collisions can become more important. In the case of compact clusters of BHs or NSs, gravitational radiation is also important, as it drives 2-body binary formation and binary mergers  \cite{Quinlan:1989,Quinlan:1990}. 

The kinetic energy released from binary formation produces heating in the core\cite{Henon:1965,Heggie:2009,McMillan:1990}, a process dubbed ``binary burning". BHs in the core may also be flung into wide eccentric orbits by dynamical interactions and inject energy into the wider cluster \cite{Kremer:2019rid}. A central SMBH or IMBH (if formed) can also provides a heat source which can halt core collapse,  causing the cluster to expand and potentially even disperse \cite{Shapiro:1977, Marchant:1980,Duncan:1983,Baumgardt:2004zu}. Accordingly, the presence of a BH subsystem and/or a IMBH serves as a possible explanation for observations of a non-core-collapsed subpopulation of stellar clusters with large well-resolved cores (see e.g. \cite{Harris:1996kt}).


Direct $N$-body integration schemes (e.g. \cite{Aarseth:2003,PortegiesZwart:2000zi,Wang:2015,Wang:2020,Rantala:2023,Sedda:2024,Barber:2024,Gaete:2024ovu,Siles:2024yym}) in principle provide the most accurate models of collisional stellar dynamics, but suffer the problem that the computational time required for computing particle-particle interactions scales as up to $O(N^2)$ for direct sum methods and $O(N \ln N)$ for tree methods \cite{Dehnen:2011fj}, which can become prohibitive for $\gtrsim 10^{5}$ particles. Nonetheless, advances in parallelization have enabled simulations with up to $\sim 10^6$ stars \cite{Kamlah:2022,Barber:2024,Sedda:2024} with realistic models for stellar evolution and the cluster stellar populations. Several alternative statistical methods have also been developed to solve or mimic the Fokker-Planck equation for a weakly collisional system \cite{Dehnen:2011fj} including semi-analytic approaches (e.g. \cite{Fragione:2021nhb,Mahapatra:2022ngs,Chattopadhyay:2023pil,Kritos:2024sgd}), Monte Carlo methods (e.g. \cite{Freitag:2005yc,Hypki:2012uk,Joshi:2000,Rizzuto:2021,Rodriguez:2022,Kiroglu:2024xpc}) pioneered by H\'{e}non \cite{Henon:1971} and Spitzer \cite{Spitzer:book}, orbit averaging, and conduction fluid approximations \cite{Lynden-Bell:1980xip,Bettwieser:1985,Heggie:1989,Shapiro:2018vju}.

These simulations primarily assume Newtonian gravity, along with in some cases up to 2.5 or 3.5 post-Newtonian order corrections in the regularization schemes for few-body close encounters \cite{Mikkola:2007ip,Zevin:2018kzq,Rantala:2023}. As the full GR inspiral is not modelled directly, the mergers of compact objects must be approximated by rules imposing a merger for objects within some distance threshold, estimating the final mass, spin and recoil velocity of the merger product using fitting formulas to the results of numerical relativity (NR) binary merger simulations  \cite{Giersz:2015,Morawski:2018,Giersz:2024,Rodriguez:2017pec,Rodriguez:2022}. In general, Monte Carlo \cite{Rodriguez:2017pec,Rodriguez:2019huv,Rizzuto:2021,Kiroglu:2024xpc}, direct $N$-body \cite{Breen:2013vla,Barber:2024,Sedda:2024,Gaete:2024ovu,Siles:2024yym} and semi-analytic \cite{Fragione:2021nhb,Chattopadhyay:2023pil} studies have found that: a significant fraction of BH binaries are retained within the cluster; that mass segregation leads to a BH-dominated subcluster in the core; dynamical encounters occur between the BHs, most frequently in the core, and in some cases this can produce hierarchical mergers and IMBH formation, either through the merger of and growth of massive stars or BHs (see e.g. \cite{Rizzuto:2021,Barber:2024} for examples of IMBH formation in $N$-body simulations and \cite{Fragione:2021nhb} for a semi-analytic treatment). The ejection of BHs due to gravitational recoil and 3-body interactions was also observed (e.g. \cite{Barber:2024,Rodriguez:2022,Fragione:2021nhb}) with the rate of ejection and fraction of BHs and BH binaries retained found to depend on the natal spin distribution of BHs and the escape velocity of the cluster \cite{Rodriguez:2019huv,Mahapatra:2021hme}. As one might expect, more compact clusters with higher escape velocities are more likely to produce IMBHs \cite{Rizzuto:2021}.

\subsection{Strongly collisional compact relativistic clusters}

\label{sec:sc_clusters}

The methods described above are sufficient for non-relativistic, weakly collisional clusters. However, in a cluster that retains a BH population, it is possible that a BH subcluster may become sufficiently compact (potentially via continued mass segregation followed by the gravothermal collapse of the BHs in the core, perhaps aided by additional dissipative processes such as dark matter dynamical friction) that such approximate treatments are insufficient. 

Such highly relativistic, collisional, $N$-body clusters have largely been unexplored. Highly symmetric arrangements of up to 5 initially stationary BHs arranged in a line \cite{Jaramillo:2010tc}, and up to 20 arranged in a ring \cite{Ponce:2010fq}, have been evolved in studies of common horizon formation. Initial data for rings of up to 1000 BHs have also been obtained but not evolved in time \cite{Abrahams:1992ru}. For BHs without such symmetries, numerical relativity simulations of 3-body \cite{Lousto:2007rj,Campanelli:2007ea,Galaviz:2010mx,Ficarra:2023zjc} and 4-body \cite{Galaviz:2010mx,Heinze:2025usf} BH systems have been conducted but not for clusters with $N > 4$.

There remain several key questions that cannot be answered reliably except by fully general-relativistic $N$-body simulations. They include the following: 

\begin{enumerate}
   \item How exactly do larger BHs form via repeated mergers: through runaway growth of a single most-massive object, or via mergers evenly distributed through the cluster so that all objects grow at comparable rates? What is the rate at which mergers occur, compared to the other timescales characterizing the cluster?

   \item Does the recoil from BH mergers, or 3-body interactions, cause BHs to be ejected from the cluster? If so, with what characteristic velocity? Does this cause the cluster to disintegrate?

   \item What does the gravitational wave signal look like from a relativistic cluster of compact objects undergoing close flyby encounters and mergers? Can individual mergers and close encounters be identified above the cluster background radiation and do they show signatures of occurring in a dense cluster environment? 

   \item What are the masses and spins of the daughter BHs? How do these evolve over multiple mergers? 

   \item What is the final fate of a bound, relativistic collisional BH cluster?
\end{enumerate}

In Section \ref{sec:results} we describe how our results relate to each of these key questions.
 
\section{Order of magnitude estimates}
\label{sec:OOM}

We can estimate the timescale for the overall dynamical evolution of our relativistic cluster and the {\it initial} timescales for BH binary formation and mergers, and direct collisions. Estimating these
timescales is useful so that we may anticipate if and when these events are likely to be triggered in our simulations. It will become evident below that there are just two parameters, $N$ and $R/M$, that essentially determine the magnitude of all of these timescales.

\subsection {Dynamical Timescale}

As we discuss in Sec. \ref{sec:sim_setup}, our simulation uses initial data based on a collisionless, monoenergetic cluster, where ``monoenergetic" means all particles have the same energy at infinity. The virial theorem gives the velocity dispersion as $v^2 \sim M/R$ up to a factor of order unity. For the collisionless monoenergetic cluster the half-mass radius is $R_h \sim 0.7 R$.  Using $v^2 \sim 0.4 M/R_h$ from \cite{Spitzer:1971} Eqn. (2) and $R_h \sim 0.7 R$ gives $v^2 \sim 0.6 M/R$. Alternatively, we can approximate the monoenergetic cluster as a $\Gamma = 3$ polytrope (true in the Newtonian limit \cite{Shapiro:1985b}) with
\begin{equation}
    2 T+W = 0 = 2 \frac{1}{2}M v^2  - \frac{3(\Gamma -1)}{(5\Gamma -6)}\frac{M^2}{R},   
\end{equation}
where $T$ is the total cluster kinetic energy and $W$ the gravitational potential energy, giving $v^2 \sim \frac{2}{3} M/R$, roughly the same result. Using $v^2 \sim \frac{2}{3} M/R$ hereafter, Eqn. \eqref{tdyn} yields
\begin{equation}
    t_{\textup{dyn}}/M \approx 30 \left(\frac{R}{10M}\right)^{3/2}.
\end{equation}

\subsection{Binary Formation and Merger timescales}

Forming a bound binary system from two objects in a cluster requires a mechanism to remove energy from the orbit. For BHs there are two main ways this can occur \cite{Quinlan:1987,Quinlan:1989}: nondissipative 3-body encounters, where a third body carries away the excess kinetic energy, and 2-body encounters, with a dissipative loss of energy via gravitational radiation. For stars there is also the possibility of dissipation via tidal effects (see \cite{Spitzer:book} for a review and further references). First, let us consider binaries formed from 3-body interactions. These tend to be ``soft" on formation, and so most end up disrupted via interactions with other stars \cite{Heggie:1975,Quinlan:1989}. The formation rate for so-called 3-body ``immortal" binaries (ones that are not disrupted and become hard) per unit volume per unit time for a steady-state equilibrium cluster of equal-mass stars with a Maxwell-Boltzmann velocity distribution is estimated to be \cite{Goodman:1985,Ginat:2024npu}
\begin{equation}
    \dot{n}_{\textup{B3}} = \frac{126 m^5 n^3}{v^9},
\end{equation}
where $n$ is the number density. From this we can get the formation timescale as 
\begin{align}
    t_{\textup{B3}} =& \frac{n}{\dot{n}_{\textup{B3}}} = \frac{v^9}{126 m^5 n^2}, \\
    t_{\textup{B3}}/M \sim& \; 1 \times 10^4 \left(\frac{R}{10M}\right)^{3/2}\left(\frac{N}{25}\right)^3,
\end{align}
setting $n = N/(4\pi R^3/3)$, $v^2 = \tfrac{2}{3} M/R$ and $m = M/N$.

For binaries formed via gravitational radiation the maximum periastron separation $r_{p,\mathrm{max}}$ can be obtained by equating the energy radiated along a parabolic orbit with periastron $r_{p,\mathrm{max}}$ (Eqn. (15) in \cite{Quinlan:1987}) with the typical binary kinetic energy in the center of mass frame $\mu \left\langle  v^2_{rel} \right\rangle/2 = m v^2/2$ where $\mu = \frac{m_1 m_2}{m_1+m_2} = m/2$ is the reduced mass and $\left\langle  v^2_{rel} \right\rangle = 2 v^2$ the mean square relative velocity. This gives (see Eqn. (16) in \cite{Quinlan:1987})
\begin{equation}
    r_{p,\mathrm{max}}/m \sim 6 \left(\frac{R}{10M}\right)^{2/7}.
\end{equation} 
Allowing for gravitational focusing, the capture cross section is then (see \cite{Quinlan:1987}, Eqns. (14)-(19))
\begin{align}
    \sigma_{\textup{cap}} \sim &\; 6\pi m^2 v^{-18/7}\left[1 + \frac{3}{2}v^{10/7}\right], \\
    \sim & \;6 \times 10^2 m^2 \left(\frac{R}{10M}\right)^{9/7} \left[1 + 0.2 \left(\frac{R}{10M}\right)^{-5/7}\right].
\end{align}
Neglecting the term in square brackets the 2-body radiation-driven binary formation timescale is then 
\begin{equation}
    t_{\textup{B2}} \sim  \frac{1}{n \sigma_{\textup{cap}} v} \sim \frac{v^{11/7}}{6\pi m^2 n},
\end{equation}
or
\begin{equation}
    t_{\textup{B2}}/M \sim \;7 \times 10^{2} \left(\frac{R}{10M}\right)^{31/14}\left(\frac{N}{25}\right).
\end{equation}
The ratio of these two timescales is 
\begin{equation}
    \frac{t_{\textup{B3}}}{t_{\textup{B2}}} \sim 17 \left(\frac{R}{10M}\right)^{-5/7}\left(\frac{N}{25}\right)^2.
\end{equation} 
For our simulations with $R \sim 10M$ and $N = 25$ we therefore expect 2-body binary capture to dominate over 3-body binary formation. Now, $t_{\textup{B3}}, t_{\textup{B2}}$ are the typical formation timescales for an individual binary. With $N > 1$ BHs in the cluster to choose from, the average times until the \textit{first} binary forms will be smaller, given by 
\begin{equation}
    t_{\textup{B3,first}}/M \sim\; \frac{3}{N} t_{\textup{B3}}/M \sim 1 \times 10^3 \left(\frac{R}{10M}\right)^{3/2}\left(\frac{N}{25}\right)^2,
\end{equation}
and
\begin{equation}
    t_{\textup{B2,first}}/M \sim\; \frac{2}{N} t_{\textup{B2}}/M \sim 50 \left(\frac{R}{10M}\right)^{31/14},
\end{equation}
where the factors of 3 and 2 are to avoid overcounting.

Next we consider the typical merger timescales for binaries formed by both mechanisms. The energy radiated in gravitational waves per unit time can be estimated from the quadrupole formula, giving a merger timescale for an equal-mass binary \cite{Peters:1964,Zwick:2019yjl} of 
\begin{align}
    t_{m} \sim & \frac{5 a_0^4}{512 m^3} f(e_0)^{-1}, \\
       f(e) =& (1-e^2)^{-7/2}\left(1 + \frac{73}{24}e^2 + \frac{37}{96}e^4\right),   
\end{align}
where $a_0$ is the initial semi-major axis and $e_0$ the initial eccentricity. We can relate $a_0$ to the the dispersion velocity $v$ and initial binary hardness $x \equiv \epsilon_B/(mv^2/3)$ \cite{Quinlan:1987}, where $\epsilon_B$ is the binding energy of the binary, as 
\begin{equation}
    a_0 = \frac{3 m}{2 x v^2},
\end{equation}
which then gives
\begin{equation}
    t_{m} \sim \frac{405 m}{8192 v^8x^4 } \frac{1}{f(e_0)}. 
\end{equation}
As $f(e_0)$ is a strictly increasing function of $e_0$, we can get a rough upper bound on the merger time by considering a circular orbit with $e_0 = 0$ and marginal hardness with $x = 1$. This gives 
\begin{equation}
    t_{m}/M \lesssim 1 \times 10^2 \left(\frac{R}{10M}\right)^4 \left(\frac{N}{25}\right)^{-1}.
\end{equation}
For the compact clusters we examine below with $R \approx 10 M$ and $N = 25$ this this upper bound on $t_m$ is already shorter than $t_{\textup{B3}}$ and $t_{\textup{B2}}$, so the inspiral time will be small compared to the binary formation time by either process.
 
\subsection{Direct collision timescales}

We can consider binary mergers as occurring either from the evolution and inspiral of hard compact binaries through gravitational radiation, or through direct collisions in otherwise hyperbolic encounters. 

For the BH direct collision timescale for hyperbolic orbits we can consider several levels of approximation. First, we can consider the BHs as collisional point-particles with equal mass $m$ and a collision cross-section $\sigma = \pi b^2$ determined by a critical impact parameter $b$. For a given collision velocity $v$ we can estimate $b$ using conservation of angular momentum for a grazing collision as 
\begin{equation}
    v b = 2 r_* v_*
\end{equation}
and take $r_* \sim m = M/N$ and $v_* \sim c \equiv 1$ giving $\sigma \sim 4\pi m^2 / v^2$. The collision timescale is then
\begin{equation}
    t_{\textup{coll}} \sim \frac{1}{n \sigma v}.
\end{equation}
Putting this together gives
\begin{equation}
    t_{\textup{coll}}/M \sim 2 \times 10^3 \left(\frac{R}{10M}\right)^{5/2} \left(\frac{N}{25}\right).
\end{equation}

A slightly improved estimate can be obtained by using the capture cross-section for a BH and a non-relativistic ($v\ll 1$) point-mass particle $\sigma = \pi b^2_{\textup{crit}} = \pi(4 m_{\textup{tot}}/v)^2$ (see e.g. \cite{Shapiro:1983}). If we adapt this by taking $m_{\textup{tot}} = 2m$ for a BH-BH collision we get $\sigma = 64 \pi m^2/v^2$, giving
\begin{equation}
    t_{\textup{coll}}/M \sim 1.3 \times 10^2 \left(\frac{R}{10M}\right)^{5/2}\left(\frac{N}{25}\right).
\end{equation}

The BHs, of course, are not nonrelativistic point particles. The results of numerical relativity simulations \cite{Shibata:2008rq,Sperhake:2009jz} give the critical impact parameter for merger of two BHs as 
\begin{equation}
    b \approx 2.5 m_{\textup{bin}} /v,
\end{equation}
for $0.6 \lesssim v \lesssim 0.9$ and where $m_{\rm bin}$ is the mass of the binary. Using this we obtain 
\begin{align}
\label{tcoll}
    t_{\textup{coll}}/M \sim& \;0.05 \frac{v}{M n m_{\textup{bin}}^2}, \\
    \sim& 3.4 \times 10^2 \left(\frac{R}{10M}\right)^{5/2}\left(\frac{N}{25}\right).
\end{align}
for $m_{\textup{bin}} = 2m$. Again, this is the timescale for an individual collision. The average time before the \textit{first} collision will be less than this, given by 
\begin{equation}
    t_{\textup{coll,first}}/M \sim \frac{2}{N}t_{\textup{coll}}/M \sim 28 \left(\frac{R}{10M}\right)^{5/2}. 
\end{equation}

Using Eqn.~(\ref{tcoll}) the ratio of collision time to dynamical time is then approximately 
\begin{equation}
    \frac{t_{\textup{coll}}}{t_{\textup{dyn}}}  \sim 13 \left(\frac{R}{10M}\right)\left(\frac{N}{25}\right), 
\end{equation}
confirming that more compact clusters with smaller numbers of particles are more strongly collisional. 

\begin{figure}
    \centering
    \includegraphics[width=\linewidth]{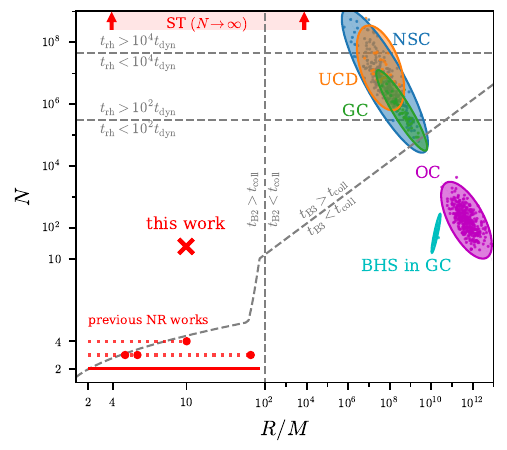}
    \caption{An $N$ vs $R/M$ diagram showing some typical astrophysical clusters, select numerical relativity (NR) studies, and transition lines for various key timescales. The astrophysical clusters are nuclear star clusters (NSC), globular clusters (GC) and ultra compact dwarf galaxies (UCD) (data from \cite{Antonini:2016gqe} Fig. 2), open clusters (OC, data from \cite{PortegiesZwart:2010cly} Fig. 2), and the estimated BH substystems (BHS) in galactic globular clusters in the MOCCA I database (\cite{Askar:2018} Table II). The numerical relativity (NR) works are \cite{Campanelli:2007ea,Galaviz:2010mx,Ficarra:2023zjc} (with approximate $R/M$ values due to the small $N$) with an additional generic line for the many NR studies of $N=2$ merging BH binaries. The collisionless (effectively $N \rightarrow \infty$) simulations of Shapiro \& Teukolsky~\cite{Shapiro:1985a,Shapiro:1985b,Shapiro:1985c,Shapiro:1986} are indicated by the red arrows and faded red rectangle, with the spacing of the arrows indicating the range of $R/M$ explored in those works.}
    \label{fig:N_vs_R_over_M}
\end{figure}

Based on these estimates the most likely mechanism to produce the first mergers for our clusters of $N = 25$, $R \sim 10M$ is direct collisions, which will occur on a timescale comparable to the dynamical timescale, followed by 2-body gravitational radiation-driven inspiral. For large enough $N$, as the BHs merge the factor $m_{\textup{bin}}^2$ for the largest BHs will increase faster than the number density, $n$, will decrease. The larger BHs will also mass segregate to the center of the cluster where $n$ is largest. As a result we should expect $t_{\textup{coll,first}}$ to become smaller for subsequent mergers with second or higher generation BHs, leading to runaway growth, provided the large BHs remain bound in the cluster and the total number of remaining bound BHs is sufficiently large ($ \gg 1$). In Fig. \ref{fig:N_vs_R_over_M} we show a $N$ vs $R/M$ diagram with crossover lines for some of the key timescales $t_{\mathrm{rh}},t_{\mathrm{dyn}},t_{\mathrm{B2}},t_{\mathrm{B3}}$ and $t_{\mathrm{coll}}$. We include the distributions for some observed and inferred astrophysical clusters, located mostly in the large $N$, low compactness, weakly collisional regime, plus numerical relativity simulations located in the very low $N$, high compactness regime as 
well as the $N \rightarrow \infty$ collisionless simulations of  Shapiro \& Teukolsky (ST), and the simulation in this work. One can see that we explore a new region of parameter space where direct collisions dominate $(t_{\mathrm{coll}} \ll t_{\mathrm{B2}},t_{\mathrm{B3}})$. 

\subsection{BH kick velocity}

As discussed in sections \ref{sec:astro_motiv} and \ref{sec:dyn_context} recoil kicks from 3-body interactions and anisotropic emission of gravitational waves during mergers can cause BHs and BH binaries to be ejected from the cluster, limiting the potential for hierarchical merger chains. Using the formulae presented in \cite{Antonini:2016gqe} (Eqns. (17)-(20)) based on the NR simulations of \cite{Gonzalez:2006md,Lousto:2009mf,Lousto:2012su}, and assuming an isotropic and uncorrelated distribution of BH spins, one finds for equal-mass BH mergers in quasicircular orbits an average (root mean square) GW-induced recoil velocity of 
\begin{equation}
    \bar{v}_{\textup{GW}} \approx v_0 \left[\chi^2_1 + \chi^2_2 + 0.023(\chi^4_1+\chi^4_2)+0.025\chi^2_1\chi^2_2\right],
\end{equation}
where $v_0 = 1.1\times 10^{-3}c = 3.2 \times 10^{2}\;\textup{km}\;\textup{s}^{-1}$ and 
$\chi_i \equiv J_i/m_{i}^2$ where $J_i$ and $m_i$ are the angular momenta and masses of the BHs. Mergers of unequal-mass binaries in general have smaller recoil \cite{Lousto:2012su}, while eccentricity can increase the recoil velocity by a factor of $\sim 25 \%$ \cite{Radia:2021hjs}. Although the merger of two equal-mass, nonspinning BHs does not produce recoil, second generation BHs formed from two equal-mass, nonspinning parents have a generic spin of $\sim 0.7$ \cite{Gammie:2003qi,Hofmann:2016yih}. For the merger of two such second generation BHs $\chi_1 = \chi_2 = 0.7$ and we get $\bar{v}_{\textup{GW}} \sim 3.2 \times 10^{2}\;\textup{km}\;\textup{s}^{-1}$. The escape velocity from a cluster scales as \cite{Antonini:2016gqe} 
\begin{equation}
    v_{\textup{esc}} \gtrsim \left(2\frac{M}{R}\right)^{1/2} \sim 1.9 \times 10^{5} \left(\frac{M}{10R}\right)^{1/2}\;\textup{km}\;\textup{s}^{-1}.
\end{equation}
For runaway growth, successive mergers of a large BH with many small companions with isotropically distributed orbital angular momenta has been found to result in a decrease in spin, with the dimensionless spin of the large BH decreasing as $\chi \sim m^{-7/3}$ in the high-spin regime and $m^{-1/2}$ in the low-spin regime (see section \ref{sec:mass_spin_evol}) where $m$ is mass of the BH mass \cite{Hughes:2002ei,Gammie:2003qi}. Thus, in general in clusters of the compactness we consider here, GW recoil alone is unlikely to result in BH ejection. However, the exchange of momentum in many-body interactions may still eject BHs from the cluster.

\subsection{Typical astrophysical timescales}
\label{sec:timescales}

So far we have left $M$ arbitrary: our simulations could equally represent a cluster of small PBHs or a cluster of SMBHs. For a cluster of $m = 0.05 M_{\odot}$ PBHs we have $M \sim 6 (N/25) \mu\mathrm{s} $, for $m = 10 M_{\odot}$ stellar-origin BHs $M \sim 1.2 (N/25) \mathrm{ms}$, for a cluster of $m = 10^6 M_{\odot}$ SMBHs we have $M \sim 2(N/25)\;\mathrm{minutes}$, and for a cluster of very massive $m = 10^9 M_{\odot}$ SMBHs we have $M \sim 1.4(N/25)\;\mathrm{days}$. 

The timescales for the compact clusters with $N = 25$ we consider in this paper are thus extremely short compared to typical astrophysical or cosmological times. However, we can use the scaling relations to compute the corresponding scales for larger, less-compact systems. For this comparison it is sometimes more useful to express the timescales in terms of the mass of the individual BHs and the cluster dispersion velocity, e.g., 
\begin{align}
    t_{\textup{B3,first}} \sim&\;  10^{7}\textup{years} \left(\frac{m}{10 M_{\odot}}\right)\left(\frac{v}{100\textup{km}\;\textup{s}^{-1}}\right)^{-3}\left(\frac{N}{10^{3}}\right)^{3}, \\
    t_{\textup{B2,first}} \sim&\;  10^{6}\textup{years} \left(\frac{m}{10 M_{\odot}}\right)\left(\frac{v}{100\textup{km}\;\textup{s}^{-1}}\right)^{-\tfrac{31}{7}}\left(\frac{N}{10^{3}}\right), \\
    t_{\textup{coll,first}} \sim&\;  10^{7}\textup{years} \left(\frac{m}{10 M_{\odot}}\right)\left(\frac{v}{100\textup{km}\;\textup{s}^{-1}}\right)^{-5}\left(\frac{N}{10^{3}}\right).
\end{align}
We can thus see that for less compact clusters with smaller dispersion velocity $v$ the binary formation channels will dominate over direct collisions.

\subsection{Simulation Timesteps}
\label{sec:sim_timesteps}

We now consider the relevant timescales for a feasible numerical simulation. The minimum timestep $\dd t_{\textup{min}} = C_{\textup{Courant}} \dd x_{\textup{min}}$ is set by the Courant factor $C_{\textup{Courant}}$ and the minimum grid spacing $\dd x_{\textup{min}}$, which is in turn set by the size of the BHs 
to ensure a minimum number of grid points covering the horizon. In the moving puncture gauge the apparent horizon of a BH mass $m$ is approximately $2m$ in diameter \cite{Radia:2021smk} so if we require at least
32 grid points across the horizon (see Appendix \ref{app:grids}) this gives $\dd x_{\textup{min}} \lesssim \frac{2m_0}{32} = \frac{M}{16N}$. The number of timesteps required to see one or more collisions is then
\begin{equation}
    \frac{t_{\textup{coll,first}}}{\dd t_{\textup{min}}} \gtrsim 3.7\times 10^{4}\left(\frac{C_{\textup{Courant}}}{0.3}\right)^{-1}\left(\frac{R}{M}\right)^{5/2} \left(\frac{N}{25}\right).
\end{equation} 
This scaling with $N$ and $R/M$ motivates starting with a relatively compact cluster with a modest number of BHs, i.e. $N = 25$.

\section{Numerical method}

\subsection{Simulation setup}
\label{sec:sim_setup}

We want the initial distribution of BH positions and momenta to approximate a cluster in virial equilibrium and stable. Our approach is to make use of the results obtained in \cite{Shapiro:1985a,Shapiro:1985b} for the evolution of relativistic collisionless clusters. We reconstruct an equilibrium collisionless cluster which was shown to be
stable, then draw from its density and velocity distribution to assign initial values for our BHs. While, as discussed in section \ref{sec:dyn_context}, it would take an infinite number of particles of infinitesimal mass (such that the
cluster rest mass $M_0$ is finite) to perfectly reproduce the results for the selected collisionless cluster, this approach should at least give us a cluster which is bound and not prone to immediate dispersion or gravitational collapse. 

As a consequence of Jeans' theorem, a general equilibrium solution of the Vlasoff equation for a collisionless cluster in a static, spherical spacetime is provided by any particle distribution function $f(E,L,m_0)$ which is a function only of the conserved constants of motion for a particle orbit in the spacetime: the energy at infinity $E$, the angular momentum $L$ and the particle rest mass $m_0$. For a cluster of equal-mass particles we can ignore the dependence on $m_0$, and dropping the dependence on $L$ gives us an isotropic velocity distribution. There remains the choice of function $f(E)$. A particularly simple model is a monoenergetic cluster, where $f(E) \propto \delta(E - E_0)$ and where $E_0$ is a constant, and that is what we use here. Such clusters can be parameterized, up to a dimensional rescaling, by a single dimensionless parameter such as the central-to-surface redshift parameter $y_c$ defined in \cite{Shapiro:1985b}, where $y_c^{-1/2} - 1$ is the redshift of a photon emitted at the cluster center and observed at the surface and the condition $y_c - 1 \ll 1$ corresponds to the Newtonian limit. It was found numerically that for a monoenergetic equilibrium cluster to be stable against collapse to a BH $y_c$ must be greater than some critical value, corresponding to a point near the maximum in the binding energy along the equilibrium sequence. This in turn corresponds to a maximum compactness $M/R$ for stability. We choose a value of $y_c = 0.819$, which corresponds to the most compact of the stable clusters evolved in \cite{Shapiro:1985b}, with a compactness of $R/M \approx 9.95$. The mass $M$ of the cluster is
completely arbitrary, thus the model, and our simulation, applies to all cluster masses with this BH phase space distribution and compaction.

\begin{figure}
    \centering
    \includegraphics[width=\linewidth]{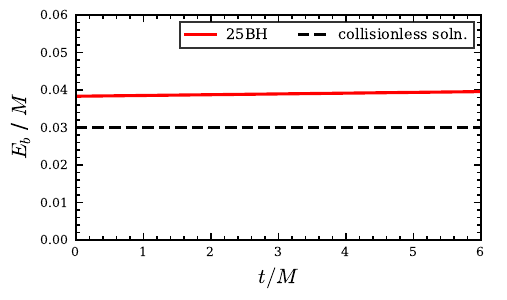}
    \caption{Ratio of the gravitational binding energy to the ADM mass, $E_b/M = (M_0 - M)/M$, vs. time  near the start of the simulation. The red line shows the numerical result and the black dashed line shows the value for the mean-field collisionless solution (corresponding to an infinite number of particles of infinitesimal mass).}
\label{fig:Eb_over_M_vs_t}
\end{figure}

We then solve the Einstein equations for this collisionless cluster, obtain its rest mass density $\rho_0(r)$ and (isotropic) velocity dispersion $v^2(r)$ profiles and randomly sample $N$ positions and $3-$velocities according to these distributions for our chosen $N$ (details of the method are given in Appendix \ref{app:construction_method}). This recipe gives us a set of initial positions and 3-momenta $\{x_i,p_i\}$ for the $N$ BHs, $i= 1 \dots N$. The idea is that for $N = \infty$ we would recover the smooth, stable, collisionless relativistic cluster we constructed above, and for finite $N$ obtain a cluster that is bound and initially in approximate equilibrium, stable against global collapse. As we are dealing with BHs, rather than point particles, we must adjust the set of initial punctures masses using an iterative routine to obtain BHs which all have the same initial horizon mass $m_{\mathrm{AH}} = m_0$. Below we settle on a final mass that is several percent smaller than what would be the particle rest mass for the corresponding mean-field collisionless cluster if sampled by only N particles. However, we verify that the resulting BH cluster is nonetheless bound and quasistable. In Fig. \ref{fig:Eb_over_M_vs_t} we show the binding energy $E_b = M_0 - M$ normalized by the ADM mass $M$. The red line shows the numerical result where the ADM mass is computed using Eqn. \eqref{eq:Mint} evaluated at $r_{\textup{ex}}/M = 50,100,133$ and extrapolated to $r_{\textup{ex}}\rightarrow \infty$ and $M_0$ is now equated to the sum of the initial BH isolated horizon masses $m_0$. The black dashed line shows the value for the smooth collisionless solution (corresponding to an infinite number of particles of infinitesimal mass). One can see the binding energy is positive, indicating the cluster is bound as desired. Its early evolution also confirms stability against global collapse.


\subsection{Initial data and grid structure}
\label{sec:initial_data}

The NR code we use is \textsc{GRChombo}, a well-tested open-source \texttt{C++} code \cite{Andrade:2021rbd,Clough:2015sqa}. \textsc{GRChombo} is particularly well-suited to this $N$-body BH problem, compared to some other NR codes, as it implements full block-structured adaptive-mesh refinement (AMR). The computational domain is covered by a set of \textit{levels} of overlapping Cartesian grids, where a level consists of a (potentially discontinuous) set of rectangular grids of the same resolution. The resolution differs by a factor of 2 between levels, such that the spatial resolution on level $l$ is $\dd x_l = \dd x_0 2^{-l}$, where each level $l>0$ is properly nested within level $l-1$. Instead of hard-coding a particular fixed grid topology in advance (such as two sets of nested boxes for a binary system as in e.g. \textsc{Carpet} \cite{Schnetter:2003rb}), we can specify some general \textit{tagging criterion} to identify regions where we want higher resolution then let the code create the required grid topology dynamically via the Berger-Rigoutsos AMR algorithm \cite{Berger:1991}, ensuring such that tagged regions are covered by higher levels (for further details see \cite{Radia:2021smk}). 

\textsc{GRChombo} can therefore simulate $N$ BHs in the same manner as it simulates a BH binary: one chooses a tagging criterion to put higher resolution near the BH horizons, and the algorithm will generate $N$ high resolution patches. In this case we use a modified version of the tagging criterion proposed in \cite{Radia:2021smk}, requiring that:

\begin{enumerate}
    \item The apparent horizons of the BHs must be covered by a minimum number of grid points.

    \item The level boundaries around the BH horizons are sufficiently spaced out.

    \item The spherical extraction surfaces for gravitational waves (and the mass and momentum surface integrals) are covered by a minimum spatial resolution. 

\end{enumerate}

The grid setup we use to achieve these requirements is detailed in Appendix \ref{app:grids}. Once we have generated the grids we then need to obtain suitable initial data. Setting the initial spins of the BHs to zero, and using maximal slicing $(K=0)$, we solve the Hamiltonian and Momentum constraints using the Conformal Transverse-Traceless (CTT) method for Bowen-York conformally flat, multiple BH initial data as described in \cite{Aurrekoetxea:2022mpw} and implemented in \textsc{GRTresna} \cite{Aurrekoetxea:2025kmm} (formerly known as the \verb|InitialConditionsSolver| module of \textsc{GRChombo}). Second order numerical convergence of the initial data is shown in Appendix \ref{app:convergence}. For Bowen-York puncture initial data the puncture mass parameter does not necessarily equal the actual mass of the BH. We approximate the event horizon by the apparent horizon, which strictly coincides only for isolated stationary BHs. We then compute the isolated horizon mass as
\begin{equation}
    m_{\mathrm{AH}} = \frac{1}{2R_s}(R_s^4 + 4 J_s ^2)^{1/2}. \label{eq:M_ah}
\end{equation}
Here $R_s$ is the areal radius of the apparent horizon given by
\begin{equation}
    R_s = (\mathcal{A}_s/4 \pi)^{1/2},
\end{equation} 
where ${\cal A}_s$ is the horizon area, and $J_s$ is the angular momentum \cite{Dreyer:2002mx,Ashtekar:2004cn} given by
\begin{equation}
    J_s = \frac{1}{8\pi}\oint_S \phi^i K_{ij} \dd S^j, \label{eq:J_ah}
\end{equation}
where $\phi^i$ is a suitable axial vector field (tangent to the horizon surface $S$ and a Killing vector field with respect to the induced metric on $S$) and $\dd S^j$ is the outward pointing surface element. Our initial BHs have $J_s = 0$ so $m_{\mathrm{AH}} = R_s/2$. In order to achieve BHs of equal initial mass $m_0$ we measure the apparent horizon masses $m^{\textup{AH}}_i$ using the \verb|ApparentHorizonFinder| of \textsc{GRChombo} and iterate using a Newton-Raphson routine to obtain values for the puncture mass parameters $m_i$ such that all BHs have the same mass to within a relative error of $10^{-4}$ (in practice only one or two iterations is required). 

\subsection{Numerical evolution}

We adopt the standard 3+1 ADM decomposition of the metric 
\begin{equation}
    \dd s^2 = -\alpha^2 \dd t^2 + \gamma_{ij}\left(\dd x^i + \beta^i \dd t\right)\left(\dd x^j + \beta^j \dd t\right),
\end{equation}
where $\alpha$ is the lapse, $\beta^i$ is the shift and $\gamma_{ij}$ the spatial metric. We use the CCZ4 formulation \cite{Alic:2013xsa} of the Einstein equations using 6th order finite difference stencils for spatial derivatives and 4th order Runge-Kutta time integration. The dynamical variables are $\chi, \tilde{\gamma}_{ij}, K, \Tilde{A}_{ij}, \Theta, \hat{\Gamma}^i$ (see \cite{Radia:2021smk} Eqns. (1)-(12) for details) and the complete set of evolution equations are given in \cite{Radia:2021smk} Eqns. (13)-(22). The gauge variables are $\alpha$ and $\beta^i$, and we use the moving puncture gauge condition \cite{Campanelli:2005dd,Baker:2005vv,Bona:1994dr} as detailed in \cite{Radia:2021smk} Eqns. (28)-(31). The evolution equations include the damping parameters $\kappa_1, \kappa_2, \kappa_3$, and we add 6th order Kreiss-Oliger dissipation. We use the standard \textsc{GRChombo} choices of $\kappa_2 = 0, \kappa_3 = 1$, while we tune $\kappa_1$ and the Kreiss-Oliger coefficient $\sigma$ to minimize the constraint violations. We find $\kappa_1 = 0.25$ and $\sigma = 1.0$ gives the best results. The moving-puncture evolution equations for the shift contain a damping parameter $\eta$ (see \cite{Radia:2021smk} Eqn. (31)) with units of inverse mass. For equal-mass binary BH simulations $\eta$ is typically set to a constant value of $\sim 2/M \sim 1/m_{\textup{BH}}$ where $M$ is the ADM mass of the spacetime and $m_{\textup{BH}}$ the mass of each of the BHs. However, for simulations with large mass ratios between the BHs a constant $\eta$ can cause instabilities and cause the simulations to fail, as $\eta$ is either too large or too small for one of the BHs \cite{Muller:2010zze}. In our simulations repeated hierarchical mergers can (and do) give rise to large BHs and thus high mass ratio mergers. We therefore adopt an adapted version of the dynamical, spatially dependent $\eta$ prescription described in \cite{Muller:2010zze} Eqn. (5), extending it to $N$ BHs (see Appendix \ref{sec:eta_prescription}). Convergence of the $\Psi_4$ gravitational waveforms with increasing numerical resolution is shown in Appendix \ref{app:convergence}.


\subsection{Diagnostics}

\begin{figure}
    \centering
    \includegraphics[width=\linewidth]{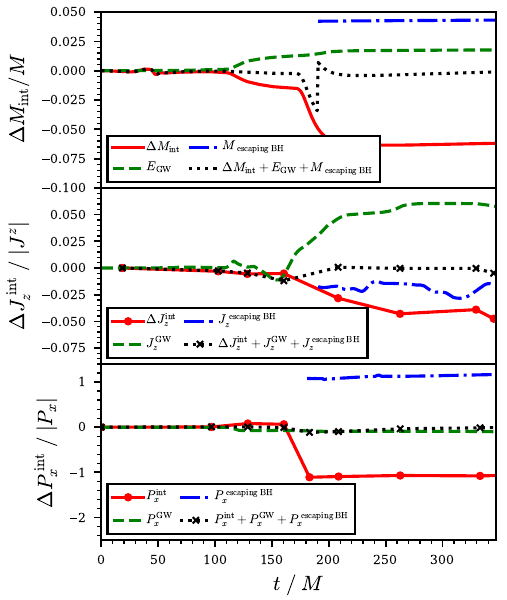}
    \caption{Conservation of mass-energy, the $z$ component of the angular momentum and the $x$ component of linear momentum. We show the change in $M_{\mathrm{int}}$, $J_z^{\mathrm{int}}$ and $P_{x}^{\mathrm{int}}$ (red solid lines), the cumulative energy and momentum lost in gravitational waves over the surface $r=r_{\mathrm{ex}}$ (green dashed lines), the mass-energy and momentum lost via the escaping BH (blue dot-dashed lines) and the sum of these quantities (black dotted lines). The integrals are evaluated over the surface at radius $r_{\mathrm{ex}}=50M$. We see that the decrease in $M_{\textup{int}}$ corresponds to the energy lost in gravitational waves, of order $\sim 1-2\%$ of $M$, the ADM mass, and the mass loss via the escaping BH. Angular momentum is conserved to within $\sim 1\%$ of the initial ADM $J_z$ ($\sim 0.18 M^2$) when the contribution from the ejected BH is included. The loss of linear momentum is dominated by the fast moving escaping BH, with the corresponding decrease in $P^{\;\mathrm{int}}_x$ matching the momentum lost via gravitational waves and the escaper to within $\sim 5\%$. Time is given in units of $M = 5(M/(10^6M_{\odot})) s$.}
\label{fig:M_ADM_cons}
\end{figure}

\begin{figure*}
\begin{tabular}{cc}
    \centering
    \includegraphics[width=0.4\linewidth]{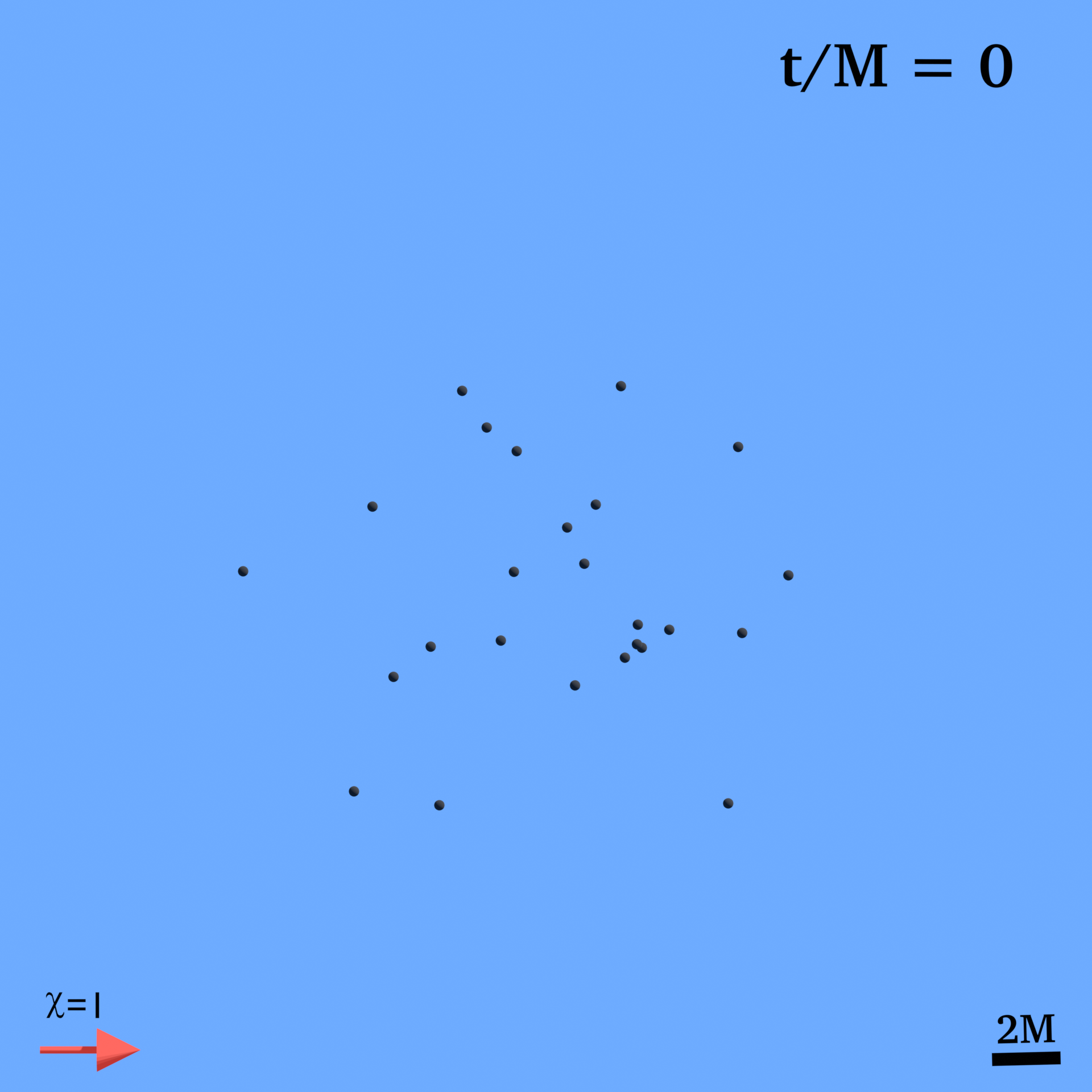} & 
    \includegraphics[width=0.4\linewidth]{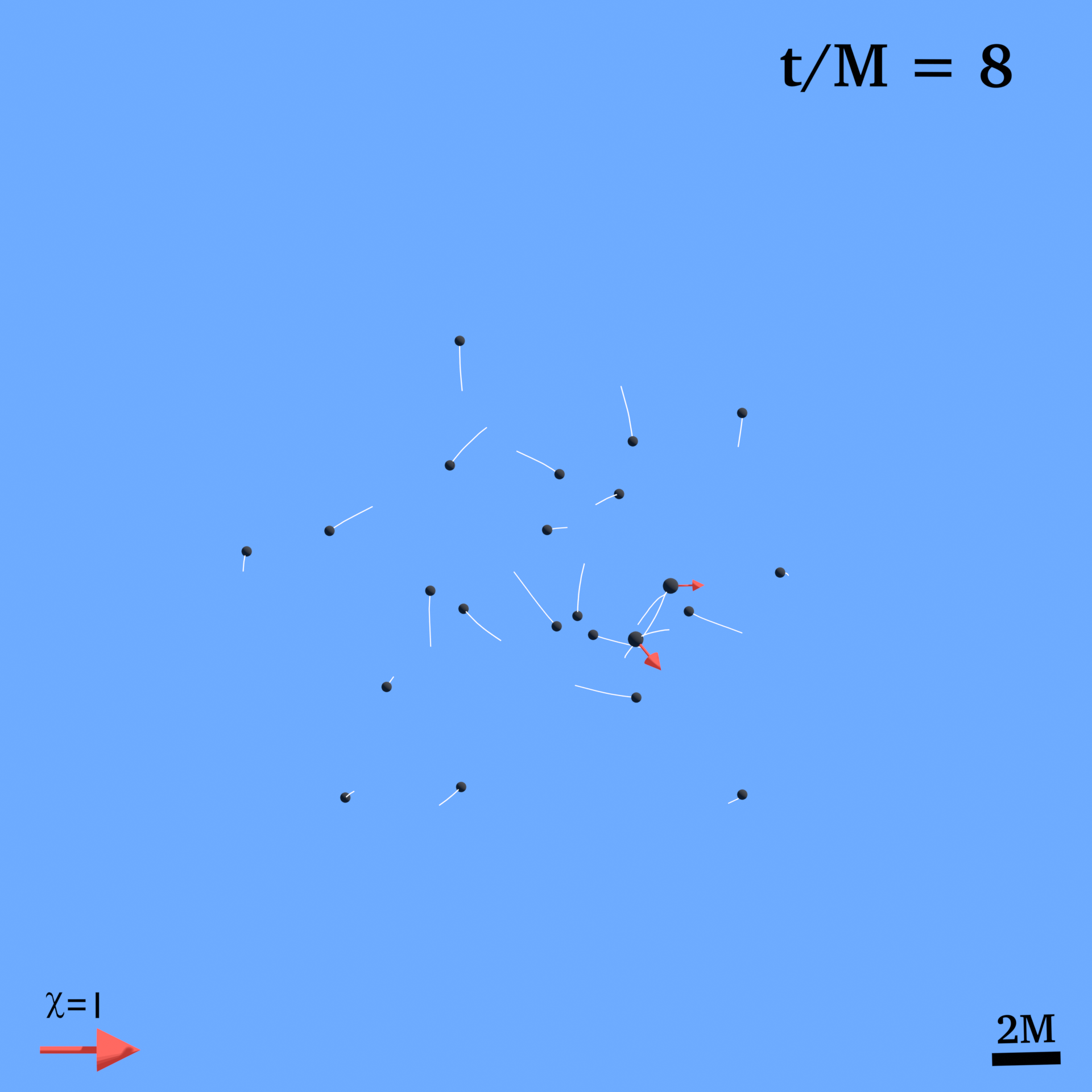} \\\includegraphics[width=0.4\linewidth]{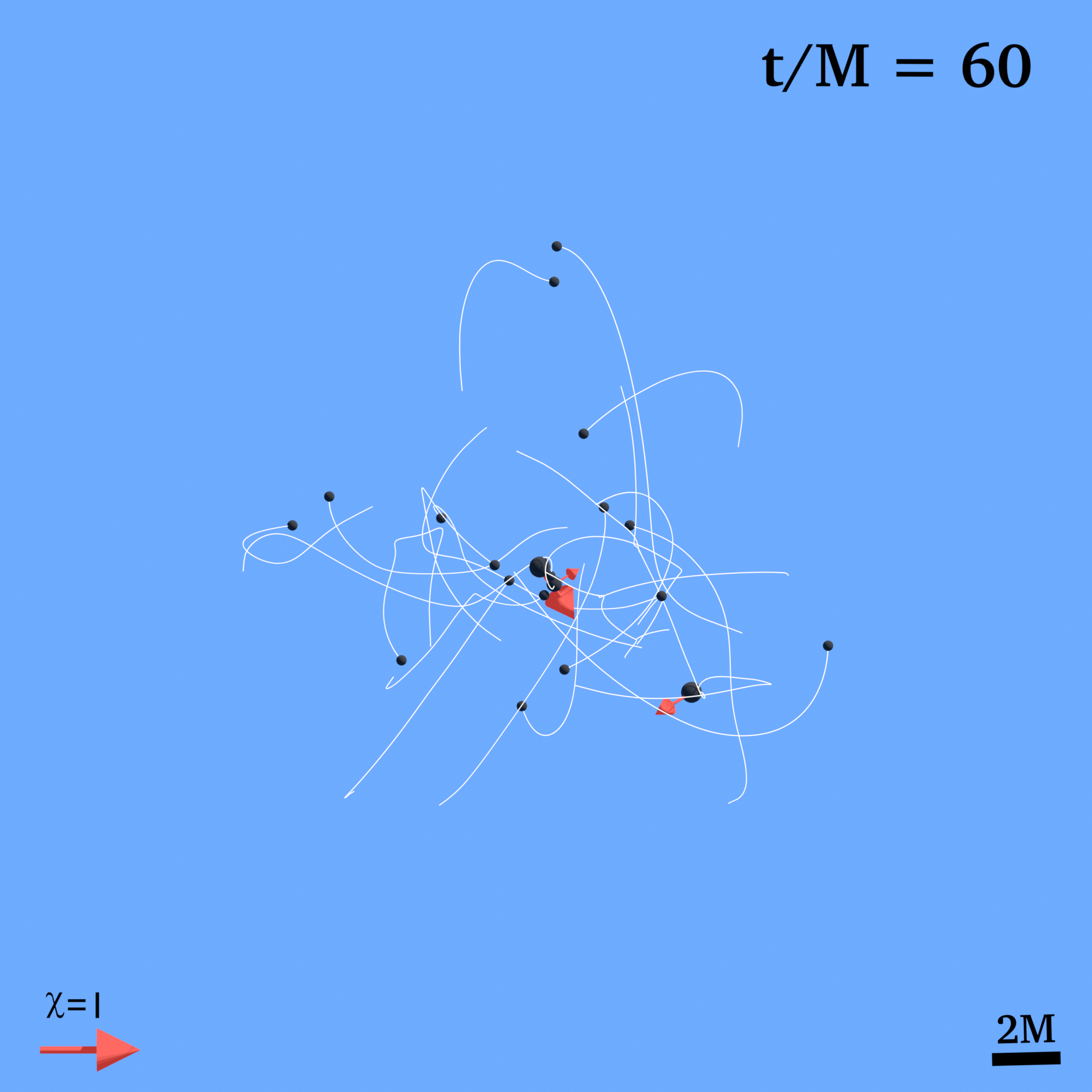} & 
    \includegraphics[width=0.4\linewidth]{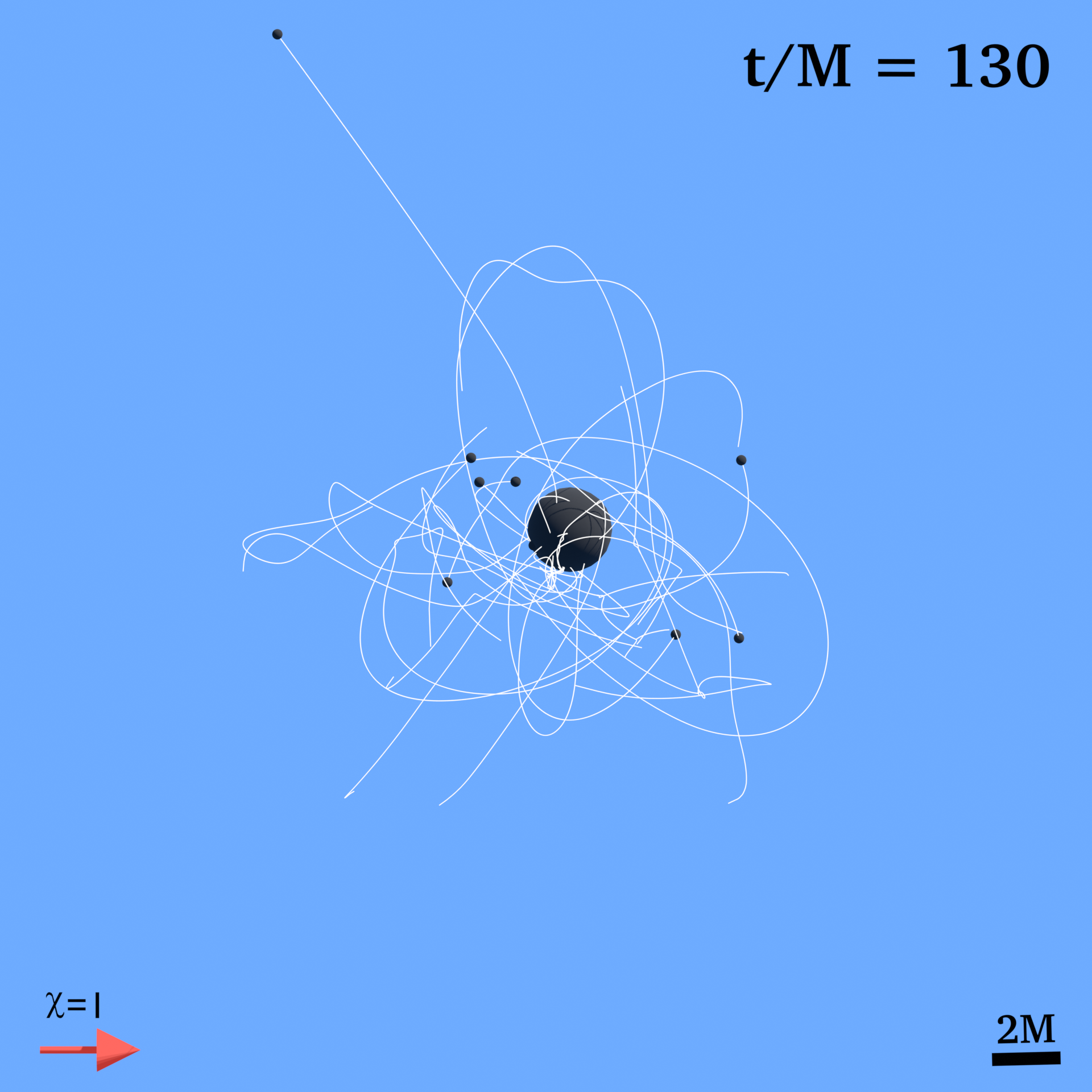} \\
    \includegraphics[width=0.4\linewidth]{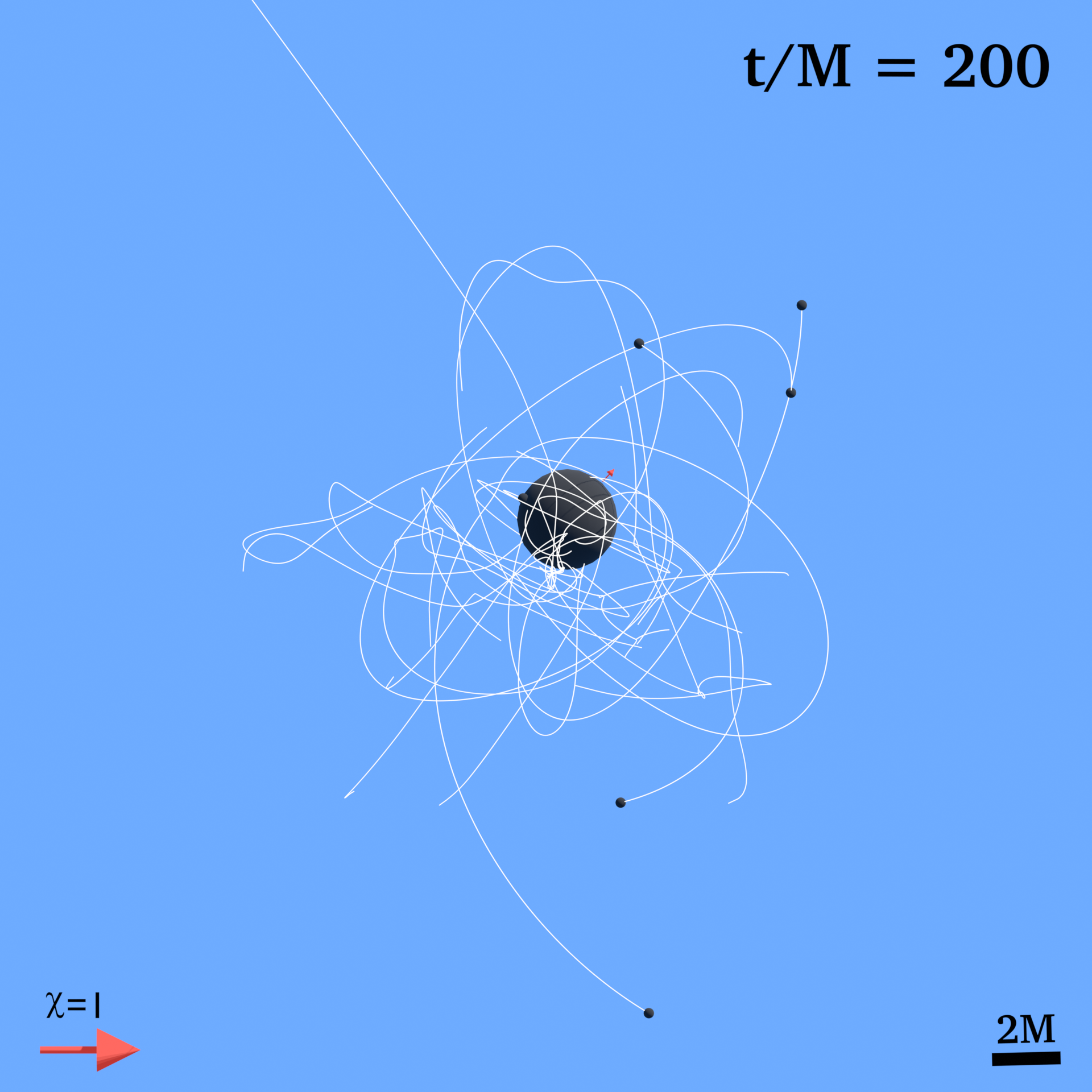} & 
    \includegraphics[width=0.4\linewidth]{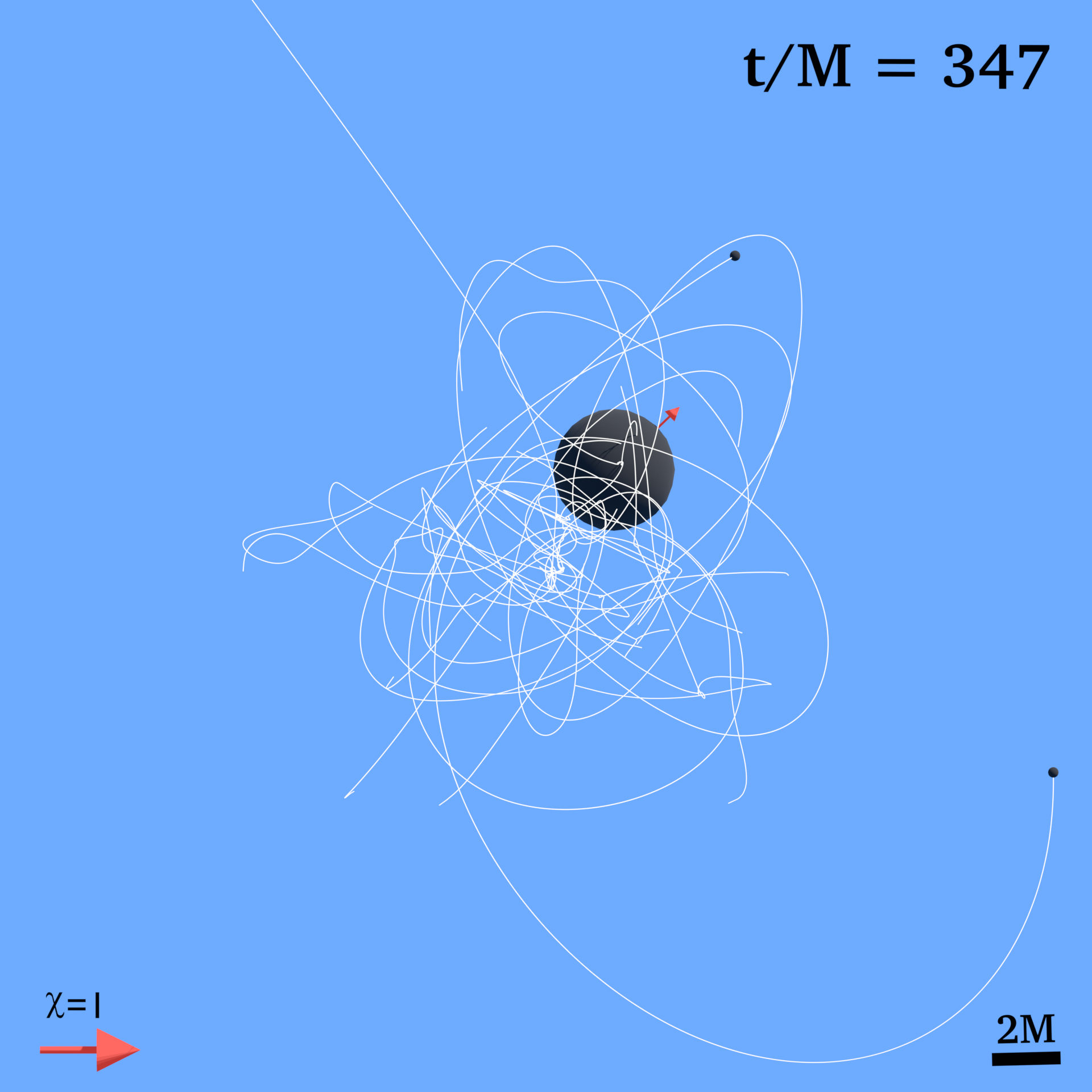}
\end{tabular}
\caption{Snapshots of the BH cluster at selected times. To make the BHs easier to see we show them as black spheres with radius $2m + const.$ where $m$ is the mass of the BH and we set the constant equal to $m_0$. The white lines show the BH puncture trajectories and the red arrows show the dimensionless spin vectors. Time is given in units of 
$M = 5(M/10^6M_{\odot}) s = 1.5\times 10^{3}(M/10^6M_{\odot}) \mathrm{km}$.}
\label{fig:3D_plots}
\end{figure*}

We extract the gravitational waves by computing surface integrals of the spherical harmonic modes of the Weyl scalar $\Psi_4$ over spherical surfaces of $r_{\textup{ex}} = 25M, 35M, 50M, 100M$ and $130M$. These can then be converted to $h_{+/\times}$ strain polarizations using the relation
\begin{equation}
    \Psi_4 = \ddot{h}_{+} - i \ddot{h}_{\times},
\end{equation}
valid for outgoing waves far from the source in the tranverse-traceless gauge. We also compute the energy and angular momentum flux radiated away in GWs (for further details see \cite{Ruiz:2007yx}), where the GW luminosity can be obtained as 
\begin{align}
    L_{\textup{GW}} \approx& \frac{r^2}{16\pi}\int \left \langle \dot{h}^2_+ + \dot{h}^2_{\times} \right \rangle \dd \Omega, \\
    \approx& \frac{r^2}{16\pi}\int \left\vert\int^t_{-\infty}\Psi_4 \dd t'\right\vert^2 \dd \Omega, \\
    \approx& \frac{r^2}{16\pi}\sum_{l,m} \left\vert\int^t_{-\infty}\Psi^{lm}_4 \dd t'\right\vert^2 \label{eq:GW_lum}
\end{align}
where $\Psi^{lm}_4$ are the spheroidal harmonic modes of $\Psi_4$ for mode numbers $l,m$, provided $r = r_{\textup{ex}}$ is sufficiently large that the surface integral is in the wave zone (see Eqns. (3.5)-(3.8) in \cite{Ruiz:2007yx}). Conservation of mass-energy, linear and angular momentum is monitored by computing surface integrals over the spheres $r = r_{\textup{ex}}$
\begin{align}
    M_{\textup{int}} =& \frac{1}{16\pi}\int  (\chi\tilde{\Gamma}^i + 2 \tilde{\gamma}^{ij}\partial_j\chi) \dd S_i, \label{eq:Mint}\\
    J_i^{\textup{int}} =& \frac{1}{8\pi}\int \epsilon_{ijk} x^j (K^{mk} - \gamma^{mk} K) \dd S_m, \\
    P_i^{\textup{int}} =& \frac{1}{8\pi}\int (K^m_i - \delta^m_i K) \dd S_m, \label{eq:P_int}
\end{align}
where $\dd S_i$ is the surface element, $\tilde{\Gamma}^i$ is the contracted conformal connection coefficient, $K_{ij}$ is the extrinsic curvature, $K$ is its trace, $x^i$ are the spatial coordinates with respect to the center of the simulation box and $\epsilon_{ijk}$ is the 3D Levi-Civita symbol. As $r_{\textup{ex}} \rightarrow \infty$ these integrals tend to the ADM mass, angular and linear momentum respectively \cite{Baumgarte:2010}. For finite $r_{\mathrm{ex}} \gg M$ we use these integrals as approximate measures for the mass-energy, angular momentum and linear momentum contained within the sphere $r = r_{\mathrm{ex}}$. We evaluate the apparent horizon masses and spins (Eqns. \eqref{eq:M_ah}-\eqref{eq:J_ah}) using the isolated horizon formalism \cite{Ashtekar:2004cn} along with a measure of the quasi-local linear momentum $P^{\mathrm{AH}}_i$ defined in \cite{Krishnan:2007pu} Eqn. (3), 
\begin{equation}
    P^{\mathrm{AH}}_i = \frac{1}{8\pi}\int_{S} (K_{jk} - \gamma_{jk} K) \xi^{j}_i \dd S^k,
\end{equation}
where $S$ is the apparent horizon surface and we use unit vectors in the Cartesian coordinate directions for $\xi^{j}_i$, so $\xi^{j}_i = \delta^j_i$, following \cite{Krishnan:2007pu}.  The ADM linear and angular momentum would be zero for an isotropic cluster of infinite infinitesimal particles; for our finite cluster of 25 BHs we find an ADM angular momentum of $J^{\mathrm{ADM}}_i/M^2 = (0.12,0.19,0.18)$ and linear momentum of $P^{\mathrm{ADM}}_i/M = (-1.9,0.10,2.2)\times 10^{-2}$. Conservation of mass-energy, angular and and linear momentum (showing $J^{\mathrm{int}}_z$ and $P^{\mathrm{int}}_x$ as examples) is shown in Fig. \ref{fig:M_ADM_cons}. Mass-energy is conserved to much better than $\sim 1 \%$, angular momentum to within $\sim 1\%$ and linear momentum to within $\sim 5\%$.


\begin{figure*}
\begin{tabular}{c}
    \centering
    \includegraphics{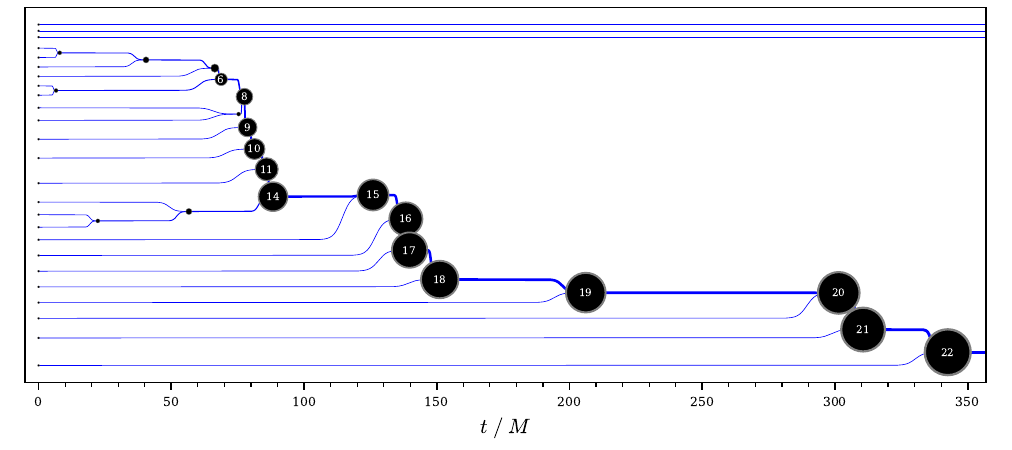}
\end{tabular}
\caption{BH merger tree. The horizontal position of the black circles show the formation times of the daughter BHs, the radius of each is proportional to the BH mass, and the white number denotes the total number of initial BHs used to form it. Time is given in units of $M = 5(M/(10^6M_{\odot})) s$.}
\label{fig:merger_trees}
\end{figure*}

\section{Results}
\label{sec:results}

Although the simulation presented here is preliminary, we can already provide tentative answers to the questions posed in Sec. \ref{sec:sc_clusters}.

\subsection{How do larger BHs form via repeated mergers? What is the rate of mergers?}
\label{subsec:merger_dynamics}

Snapshots at select times are shown in Fig. \ref{fig:3D_plots}, while the merger tree is shown in Fig. \ref{fig:merger_trees}. The initial mergers are fairly evenly distributed through the cluster, with three different pairs of first generation mergers forming three second generation BHs with mass $\sim 2m_0$ (two of these second generation BHs are evident in the upper right frame of 
Fig. \ref{fig:3D_plots}). The very first merger occurs at $t_{\textup{merger,first}}/M \approx 6$.\footnote{Two other simulations evolved with different random samples, evolved up until the first merger, give $t_{\textup{merger,first}}/M \approx 15$ and $31$ respectively.} This value is smaller than, but within an order of magnitude of, the $t_{\textup{coll,first}}/M \sim 28$ timescale estimated for direct collisions in Sec. \ref{sec:OOM}, and indeed we observe that the mergers occur via direct 2-body collisions, with very little inspiral (much less than one orbit). Note that this is very different from less compact, weakly collisional clusters. For instance, in the $N \geq 10^5$, $R/(10M) \gtrsim 2\times 10^6$ post-Newtonian globular cluster simulations of \cite{Rodriguez:2017pec} most binary mergers are quasicircular, with only $\sim 6 \%$ having eccentricity of $e \gtrsim 0.05$.  

Between $t/M \sim 40$ and $t/M \sim 60$ two of these second generation BHs each merge with another first generation BH to form two third generation BHs of mass $\sim 3m_0$ (visable in Fig. \ref{fig:3D_plots} middle row, left plot). From $t/M \sim 65 M_{\odot}$ onwards one of these BHs merges with another first generation BH, then undergoes runaway growth with a rapid sequence of mergers until it dominates the cluster (see the large central BH in Fig. \ref{fig:3D_plots}, bottom row and the right-hand plot on the middle row). This behaviour is consistent with what we predicted in section \ref{sec:timescales}: the time between mergers decreases as the most massive BH grows, then increases again as the number of BHs in the cluster is depleted through mergers and BH escape. In the bottom right plot of Fig. \ref{fig:3D_plots} at $t/M = 347$ there are only two small BHs left in bound orbits around the large, central BH. Hence by this time 22 of the initial first generation BHs have merged together to form a single large BH, and one BH has escaped the cluster. 

\subsection{Do we see BHs being ejected?}

\begin{figure}
    \centering
    \includegraphics[width=\linewidth]{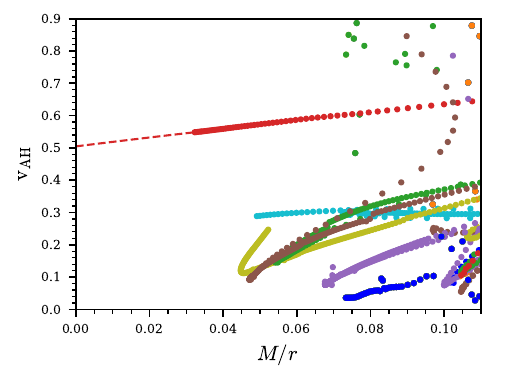}
    \caption{Apparent horizon velocity $v_{\textup{AH}}$ vs $M/r$ where $r$ is the coordinate distance to the center of the simulation box. The colored points show the numerical data for the different BHs. The red dashed line shows the analytic extrapolation of the velocity of the escaping BH (red points) to $r \rightarrow \infty$ using Eqn. \eqref{eq:Sch_v2}.}
\label{fig:sim_2_vAH_vs_R}
\end{figure}

We see clear evidence of one first generation BH being dynamically ejected from the cluster, with a substantial positive asymptotic velocity $v_{\infty} \sim 0.51c$. One can see its trajectory, moving away from the cluster in the 11 o'clock position, on the last three plots of Fig. \ref{fig:3D_plots}. To obtain $v_{\infty}$ we approximate the spacetime exterior to the cluster as a fixed Schwarzschild background metric with mass $M_*$ and approximate the single escaping BH as a point particle on a radial geodesic trajectory. The magnitude of its locally measured three-velocity should then satisfy 
\begin{equation}
    v^2(r) \approx v^2_{\infty} + (1-v^2_{\infty})2 M_*/r,
    \label{eq:Sch_v2}
\end{equation}
where $M_* \approx M - m_{\textup{0}}$ is the gravitational mass enclosed within radius $r$, excluding the escaping BH itself and neglecting any energy previously lost in gravitational waves, which is only $\sim$ 0.015M (see Fig. \ref{fig:M_ADM_cons} and Appendix \ref{sec:escaping_BH_diagnostic} for a derivation). In Fig. \ref{fig:sim_2_vAH_vs_R} we plot the velocity $v_{\textup{AH}}$ vs $M/r$ for the outermost BHs, where $v_{\textup{AH}}$ is computed from the quasi-local apparent horizon linear 3-momentum $P^{\textup{AH}}_i$ of the BHs via $\vert P^{\textup{AH}}\vert \approx m_{\textup{AH}} v_{\textup{AH}} / \sqrt{1 - v^2_{\textup{AH}}}$, and $r$ is the distance from the center of the simulation grid. We can then use Eqn. \ref{eq:Sch_v2} to obtain the asymptotic velocity $v_{\infty}$. Examination of the trajectories of the BHs suggests the cause of its ejection is a close dynamical interaction with one first generation BH, one second generation BH, and one third generation BH (the BHs with the red spin arrows visible in the center of the lefthand plot on the
second row of Fig. \ref{fig:3D_plots}) which all merge together in the course of the interaction, producing the BH that undergoes runaway growth, at $t/M \sim 65-70$. 

The escaper thus helps to drive mergers and massive BH formation by carrying away excess energy and linear momentum. We also note that $v_{\infty} \sim 0.51c$ is much larger than the maximum possible recoil velocity obtainable in a binary merger, which is roughly $28,000$km s$^{-1}$ or $0.095c$ \cite{Healy:2022jbh}. 
    
\begin{figure}
    \centering
    \includegraphics[width=\linewidth]{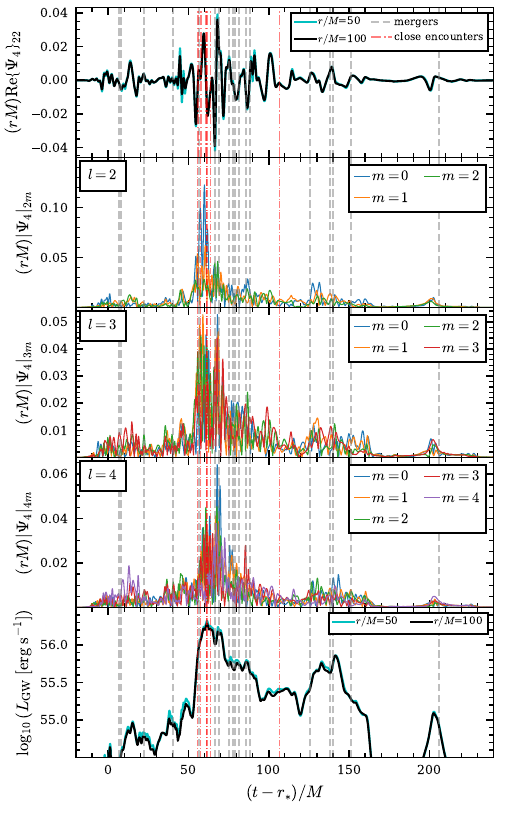}
    \caption{The gravitational wave signal, as encoded by the $\Psi_4$ Weyl scalar. The top row shows the real part of the $l=m=2$ mode, extracted at $r/M=50$ and $r/M=100$. The middle three rows show the amplitude of absolute magnitude of the $l=2,3,4$ modes vs. time, with the different colored solid lines showing different azimuthal $m$ modes from $m=0$ to $m=l$. The bottom row shows the overall luminosity, summing over all the modes. The vertical grey dashed lines show times of BH mergers, and the faded red vertical dot-dashed lines show the times of close encounters, arbitrarily defined as occasions where two BHs reach a minimum coordinate separation of $\leq 4(m_1 + m_2)$, where $m_1,m_2$ are the two BH masses, without merging. On the $x$-axis we show retarded time in units of $M = 5(M/(10^6M_{\odot})) s$.}
\label{fig:GW_signal}
\end{figure}
    
\subsection{What are the gravitational wave signals?}

The waveform of the real part of the $l=m=2$ mode of the Weyl $\Psi_4$ scalar, related to the gravitational wave tranverse-traceless strain polarizations by $\Psi_4 = \ddot{h}_+ - i \ddot{h}_{\times}$, is shown in the top row of Fig. \ref{fig:GW_signal} (note that here $m$ refers to the spheroidal harmonic azimuthal mode number, not the BH mass). On the $x$ axis we show retarded time $t - r_*$ where $r_* = r_s + 2M \ln(r_s/(2M) - 1)$ and $r_s$ is the Schwarzschild (areal) radius. The coordinate times of BH mergers and close hyperbolic and parabolic encounters, arbitrarily defined as two BHs approaching to within $4(m_1 + m_2)$ of each other without merging, where $m_1,m_2$ are the masses of the two BHs, are shown with grey dashed and red dot-dashed vertical lines respectively. In the middle three rows of Fig. \ref{fig:GW_signal} we show the amplitudes of the $l=2,3,4$ and $m= 0 \dots l$ modes. In the bottom row we show the total gravitational wave luminosity (note that this is fully independent of $M$). 

\begin{figure} 
    \centering
    \includegraphics[width=\linewidth]{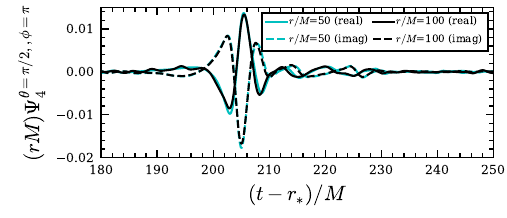}
    \caption{The gravitational wave signal, as encoded by the $\Psi_4$ Weyl scalar, viewed from the direction $\theta = \pi/2,\phi = \pi$, for the merger at $t/M \sim 206$. Solid lines denote the real part of $\Psi_4$, dashed lines the imaginary part.}
\label{fig:GW_signal_206M}
\end{figure}

There are several obvious differences from both the familiar waveform of a typical quasi-circular BH binary inspiral/merger and from the 3- and 4-body mergers studied in \cite{Campanelli:2007ea,Galaviz:2010mx,Ficarra:2023zjc}. First, as discussed in section \ref{subsec:merger_dynamics}, the compact nature of our systems means that the mergers we see are direct collisions or highly eccentric, so it is the $m=0$ modes which have the largest amplitude, rather than $m=2$. Second, during the first $\sim 150 M$ the signals of many different mergers overlap, such that one cannot easily isolate the waveform from a single merger event or single 2-body interaction. Note also that as some merger and close encounter events occur away from the origin, and the spacetime within the cluster is not a mass $M$ Schwarzschild spacetime, the travel time from an event to the extraction radius $r$ may not be exactly  $r_*$, with a difference of up to circa $\pm R \sim \pm 10M$, so the vertical lines do not necessarily all exactly line up with peaks in the waveform. Nonetheless, one can see that the largest amplitude and largest gravitational wave luminosity arise between $(t-r_{*})/M \sim 55-90$, where the largest number of mergers and close encounters occur. At late times, when the number of BHs in the cluster has been greatly depleted, the time between mergers is sufficiently large that one can distinguish individual events. A close up of the $\Psi_4$ waveform corresponding to the late-time merger at $t/M \sim 206$, viewed from one direction in the equatorial plane, is shown in Fig. \ref{fig:GW_signal_206M}. As this merger is fairly eccentric ($e \gtrsim 0.2$, see Fig. \ref{fig:final orbits}) we do not see much of the inspiral, however one can see the burst associated with the merger and the subsequent exponentially damped ringdown of the daughter BH. The ringdown signal is not especially clean as the large $q \sim 18$ mass ratio makes the merger highly asymmetric and so excites several different oscillation modes \cite{Abedi:2023kot}. In addition, the daughter BH has a non-zero velocity, the merger occurs $\sim 5M$ away from the center of the simulation box which is the center of the spherical surfaces on which we extract $\Psi_4$, and there are five other remaining BHs in bound orbits around the large daughter BH. Despite these complications we found that the measured oscillation frequencies and damping times of the dominant $l = 2$ ringdown modes were broadly consistent with the calculated values for mass and spin of the daughter BH.

\begin{figure*}
    \centering
    \includegraphics[width=\linewidth]{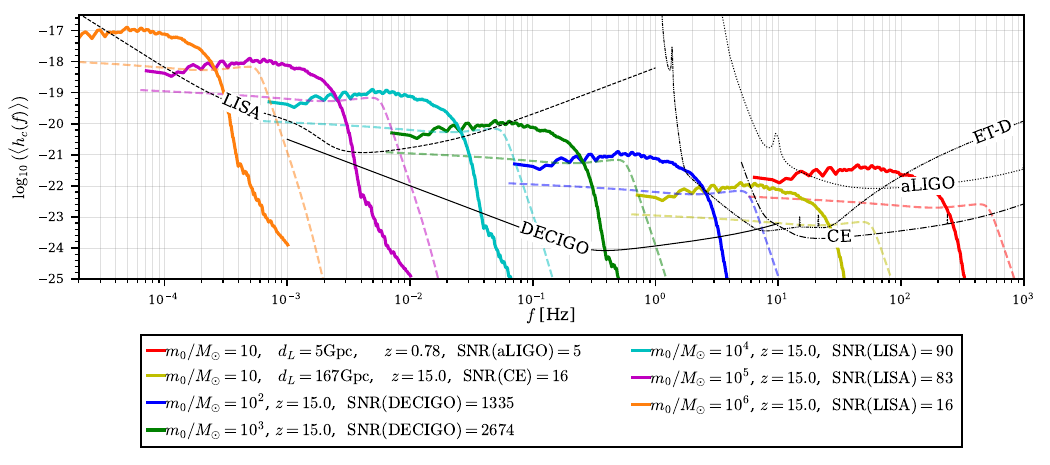}
    \caption{Characteristic strain curves (solid colored lines) for the gravitational wave signal, averaged over polarization and source orientation and scaled to different initial BH masses $m_0$, luminosity distance $d_L$ and corresponding redshift $z$ (shown in the figure caption). Also shown are the sensitivity curves (black) for different gravitational wave observatories: Advanced LIGO \cite{aLIGO:2020wna}, the Einstein Telescope (D configuration) \cite{Hild:2010id}, Cosmic Explorer \cite{Reitze:2019iox,CEcurve}, DECIGO \cite{Sato:2017dkf} (using the approximate analytic model in \cite{Yagi:2011wg} Eqn. (5)) and LISA \cite{LISA:2017} (using \cite{Robson:2018ifk} Eqn. (1)). The colored dashed lines show characteristic strain curves for a binary merger of two equal mass, nonspinning BHs of the same initial BH mass $m_0$ and source luminosity distance computed using the \textsc{IMRPhenomD} phenomological model \cite{Khan:2015jqa,phenompy}. The SNR shown in the figure caption refers to the maximum SNR for an optimum matched Wiener filter using the detector shown in parentheses. The half-mass radius of the cluster in physical units is $R_h \sim 0.7R \sim 260 (m_0 / M_{\odot}) \mathrm{km} = 10^7 (M / 10^{6} M_{\odot}) \mathrm{km}$.}
\label{fig:GW_spectrum}
\end{figure*}

\begin{figure}
    \centering
    \includegraphics[width=\linewidth]{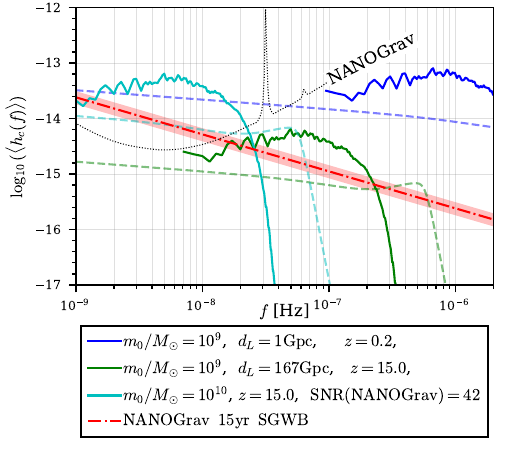}
    \caption{Characteristic strain curves for clusters of very massive SMBHs and the single-source sensitivity curve for NANOGrav \cite{McLaughlin:2013} based on the 15yr dataset \cite{NANOGrav:2023ctt} and the results of \cite{Hazboun:2019vhv}. The stochastic gravitational wave background, as measured by NANOGrav \cite{NANOGrav:2023gor}, is shown with a red dot-dashed line using the fiducial $h_c \propto f^{-2/3}$ spectrum, with the faded red rectangle showing the 90\% confidence interval. As in Fig. \ref{fig:GW_spectrum} the caption shows the luminosity distance, redshift, and, for the $m_0 = 10^{10}M_{\odot}$ curve (cyan) which lies above the NANOGrav sensitivity, the optimum matched-filter SNR.}
\label{fig:GW_PTA}
\end{figure}

In Fig. \ref{fig:GW_spectrum} we show the characteristic strain (averaged over polarization and source orientation) vs. frequency curves (thick colored solid lines) for the signal, scaled for different physical initial (source frame) BH masses and luminosity distances, from $m_0 = 10M_{\odot}$, appropriate for stellar-origin BHs, up to $m_0 = 10^6 M_{\odot}$, appropriate for a cluster of SMBHs, alongside the sensitivity curves for different gravitational wave observatories. The polarization-averaged characteristic strain is defined as 
\begin{align}
    h_c(f) =& 2 f \vert \tilde{h}(f) \vert, \\
    \tilde{h}(f) =& \sqrt{\frac{\vert \tilde{h}_+\vert^2 + \vert \tilde{h}_{\times}\vert^2}{2}}, 
\end{align}
where ``$\tilde{\;\;\;}$" denotes the
 Fourier transform, and we compute the average over source orientation angles $\theta,\phi$ as 
\begin{equation}
    \langle h_c(f) \rangle = \frac{1}{4\pi}\int \vert h_c(f,\theta,\phi) \vert \dd \Omega.
\end{equation}
The detector sensitivity curves show $h_n(f) = \sqrt{f S_n(f)}$ where $S_n(f)$ is the noise power spectral density \cite{Moore:2014lga}. We show one curve (red) for a source made up of stellar-origin initial $m_0 = 10M_{\odot}$ BHs located at a luminosity distance of $5\mathrm{Gpc}$, approximately the distance of the furthest source detected by LIGO to date \cite{LIGOScientific:2020ufj}, and six more curves for sources with $m_0 = 10$ to $10^{6}$ at a large redshift $z=15$, appropriate for BH clusters in the early universe which may help provide the seeds for  SMBH formation \cite{Kritos:2024sgd}, detectable by future space-based observatories LISA and DECIGO. The frequency of the signals is adjusted for redshift, calculated using the \verb|Planck18| model from \verb| astropy | \cite{The_Astropy_Collaboration_2022} based on the Planck (2018) parameters \cite{Planck:2018vyg}. For comparison, we also show curves (colored dashed lines) for equal-mass nonspinning BH binary mergers with the same BH mass $m_0$ and luminosity distance computed using the \textsc{IMRPhenomD} \cite{Khan:2015jqa,phenompy} model. When compared to the equivalent BH binary curves one can see that the characteristic strain spectrum for the 25 BH cluster is both significantly louder, and at a lower frequency, than the equivalent BH binary merger curve, with the largest $\langle h_c(f) \rangle$ for the cluster signal occuring at $f \sim 0.12/M$. 

In the figure caption we show the signal to noise ratio (SNR) for each curve using the detector given in parentheses, assuming an optimum Weiner filter with a perfectly matched template. This can be computed via
\begin{equation}
   \textup{SNR}^2 = \int^{\infty}_{-\infty} \left[\frac{h_c(f)}{h_n(f)}\right]^2 \dd \ln f, 
\end{equation}
(\cite{Moore:2014lga} Eqn. (19)). While this SNR is well defined and a useful guide to detectability, one should bear in mind that $N$-body systems are highly chaotic with small changes in the initial parameters leading to very different orbits and thus waveforms \cite{Galaviz:2010mx}, and it is likely our BH systems are no different. It is therefore unlikely that one could find a perfectly matched template and optimum Weiner filter for a given real cluster signal. A more realistic detection method would be a form of burst search for unmodelled or only partially-modelled signals (see e.g. \cite{Powell:2024bnp,KAGRA:2021tnv,Knee:2024mst}), resulting in a somewhat lower SNR, although a detailed analysis using realistic detector burst-search pipelines is beyond the scope of this preliminary study. Nonetheless, the $> 10$ (matched template) SNR values available for many different mass scales, even at redshift $15$, suggests that future observatories should be able to detect these systems across most of the observable universe. The detector with the largest reach is DECIGO \cite{Sato:2017dkf}, due to its high design sensitivity, most suited for observing clusters with initial BH masses in the IMBH range around $10^2-10^3 M_{\odot}$. However third generation ground based detectors like Cosmic Explorer \cite{Reitze:2019iox,CEcurve} and the Einstein Telescope \cite{Punturo:2010zz}, sensitive to clusters of stellar-origin BHs, and LISA \cite{LISA:2017}, sensitive to clusters of SMBHs and massive IMBHs, also offer excellent detection prospects. In addition, compact BH clusters will likely also be associated with other, individually identifiable, BH binary mergers, occurring in outer, less-dense regions of the host stellar or galactic cluster, or in a prior, less-compact, weakly-collisional regime (see Sec. \ref{sec:weakly_coll}). They will also likely have as electromagnetic counterparts from the gaseous host stellar cluster, host galaxy, or accreting BHs, all of which may aid with source localization and detection.

Signals from clusters of extremely massive SMBHs may also fall into the frequency band of pulsar timing arrays (PTAs). In Fig. \ref{fig:GW_PTA} we show characteristic strain curves from clusters of $m_0 = 10^9 M_{\odot}$ and $m_0 = 10^{10} M_{\odot}$, as well the corresponding signals from BH binary mergers, as in Fig. \ref{fig:GW_spectrum}, in addition to the sensitivity curve for NANOGrav \cite{McLaughlin:2013} (used as a representative PTA) based on the noise model for the 15 yr dataset \cite{NANOGrav:2023ctt} and \cite{Hazboun:2019vhv}. We also show the NANOGrav measurement of the stochastic gravitational wave background \cite{NANOGrav:2023gor} (red dot-dashed line) assuming the fiducial $h_c \propto f^{-2/3}$ slope. One can see that signals from $m_0 \lesssim 10^9 M_{\odot}$ clusters are likely beyond the reach of PTAs, although it is possible that if one were to observe the cluster on a much longer timescale if would contribute a lower frequency component that would extend into the PTA regime (analogous to the early inspiral, low frequency regime of BH binary mergers). However, if compact clusters of $m_0 \geq 10^{10} M_{\odot}$ BHs existed in the early universe these may be observable with PTAs. Unresolved systems may also make a substantial contribution to the stochastic background, although to model that we require a better understanding of the GW signal from clusters of different BH number, compactness and cluster composition and their likely distribution over cosmic time. At present we leave such a calculation for a future work.

\begin{figure}
    \centering
    \includegraphics[width=\linewidth]{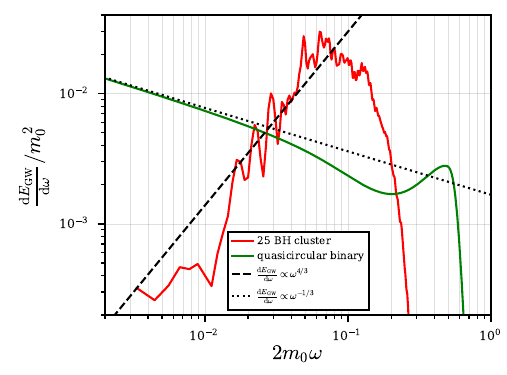}
    \caption{Gravitational wave energy per unit angular frequency $\omega=2\pi f$ normalized by $m_{0}^2$ for the 25 BH cluster (red line) and an equal-mass quasicircular binary merger with the same initial BH mass (green line), along with trendlines for $\dd E_{\mathrm{GW}}/\dd \omega \propto \omega^{4/3}$ and $\dd E_{\mathrm{GW}}/\dd \omega \propto \omega^{-1/3}$ (black dashed and dotted lines, respectively).}
\label{fig:dE_domega}
\end{figure}

From Eqn. \eqref{eq:GW_lum} the total emitted GW energy is 
\begin{align}
    E_{\mathrm{GW}} =& \int^{\infty}_{t=-\infty} L_{\mathrm{GW}}\; \dd t\\ 
    \approx& \int^{\infty}_{t=-\infty} \frac{r^2}{16\pi} \sum_{l,m}\left \vert \int^t_{-\infty} \Psi^{lm}_4(t') \dd t' \right\vert^2 \dd t, \\
    \approx& \int^{\infty}_{t=-\infty} \frac{r^2}{16\pi} \sum_{l,m} \vert H^{lm}(t) \vert^2 \dd t, 
\end{align}
evaluated at large $r$, where $H^{lm}(t) := \int^t_{-\infty} \Psi^{lm}_4(t') \dd t'$. Parseval's theorem then gives 
\begin{equation}
   E_{\mathrm{GW}} \approx \int^{\infty}_{\omega=0} \frac{r^2}{16\pi} \frac{2}{2\pi} \sum_{l,m} \vert \tilde{H}^{lm}(f) \vert^2 \dd \omega,
\end{equation}
where $\omega = 2\pi f$  and a tilde denotes the Fourier transform as before, so 
\begin{align}
    \dv{E_{\mathrm{GW}}}{\omega} \approx& \frac{r^2}{16\pi^2} \sum_{l,m} \vert \tilde{H}^{lm}(f) \vert^2, \\
    =& \frac{r^2}{16\pi^2} \sum_{l,m} \vert \tilde{\Psi}_4^{lm}(f) \vert^2 / \omega^2, \\
    =& \frac{r^2}{8\pi^2} \omega^2 \sum_{l,m} \vert \tilde{h}^{lm}(f) \vert^2, \\
    =& \frac{r^2}{8} \sum_{l,m} \vert \tilde{h}_c^{lm}(f) \vert^2.
\end{align}
In general any system which begins in a non-relativistic state with compactness $M/R \ll 1$ and small or zero initial velocities $v \ll c$, then collapses to a highly relativistic state, will produce gravitational radiation with an energy spectrum 
\begin{equation}
    \dv{E_{\textup{GW}}}{\omega}
    \propto \omega^{4/3}
\end{equation}
in the low-frequency $\omega \rightarrow 0$ limit (see \cite{Wagoner:1979} and \cite{Shapiro:1983} section 16.7 pg. 491-495). This result applies for highly eccentric ($e \sim 1$) binary mergers and for frequencies above a lower bound $\omega_* \sim (M R^{-3})^{1/2}$ set by the initial compactness of the system and below an upper cutoff set by the characteristic frequency of the relativistic phase \cite{Wagoner:1979}. In contrast, the quadrupole approximation for the energy from a quasicircular ($e \sim 0$) binary merger, valid in the low-frequency limit, gives \cite{Misner:1973prb,Shapiro:1983}
\begin{equation}
    \dv{E_{\textup{GW}}}{\omega}
    \propto \omega^{-1/3}.
\end{equation}
In Fig. \ref{fig:dE_domega} we show the gravitational wave energy spectrum from our simulated cluster and for a quasicicular binary merger with the same initial BH mass $m_{0}$, both normalized by $m_{0}^2$, with $\omega^{4/3}$ and $\omega^{-1/3}$ trendlines. On the $x$ axis we show $\omega$ normalized by $2m_{0}$, the mass of a binary. 
Even though our cluster starts in a mildly relativistic state, with $M/R \sim 0.1$, we see the spectrum still shows on average the $\omega^{4/3}$ behaviour, predicted for collapse and highly eccentric mergers, from roughly $2 m_{0}\omega \sim 0.015$ to $0.045$, with the decrease in energy at lower frequencies likely due to the finite initial size. The upper limit of $2 m_{0}\omega \sim 0.045$ is very close to that found for eccentric 2-body captures in \cite{Wagoner:1979}. The low-frequency energy spectrum may therefore serve as a useful diagnostic to distinguish eccentric mergers in compact many-body clusters from isolated quasicircular mergers if the parameters of individual merger events cannot be reliably determined.  

\begin{figure}
    \centering
    \includegraphics[width=\linewidth]{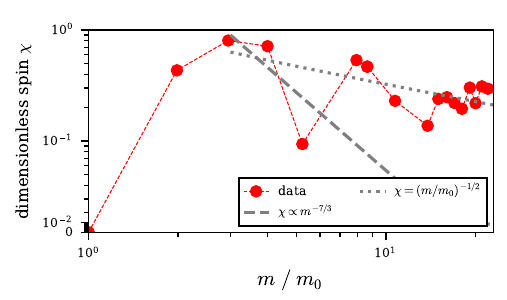}
    \caption{Dimensionless spin $\chi$ vs. BH mass $m$, normalized by the initial BH mass $m_0$, for the main (runaway) merger chain of BHs (red circles with dashed lines). We also show trendlines for spin $\propto m^{-7/3}$ and $\sim q^{1/2} \sim (m/m_0)^{-1/2}$, the early and late time relationships described in the model of \cite{Hughes:2002ei,Gammie:2003qi} for a central BH slowly accreting many smaller objects with an isotropic orbital angular momentum distribution.}
\label{fig:spin_vs_mass}
\end{figure}

\subsection{How do the masses and spins of the BHs evolve over many mergers?}
\label{sec:mass_spin_evol}

As discussed in section \ref{sec:initial_data} our cluster is initially constructed of equal-mass BHs, with zero spin and mass $m_0$. The second generation of BHs, formed by mergers of these initial spinless BHs, obtain dimensionless spins of between $\chi \sim 0.15$ and $\chi \sim 0.7$. The lower end of this range range is smaller than the typical $\chi \sim 0.7$ expected for equal-mass quasicircular binary mergers \cite{Berti:2007fi},\cite{Gammie:2003qi},\cite{Rodriguez:2019huv}. This is likely a consequence of the highly eccentric collision-like nature of the mergers. Although the spin of the daughter BH has been found to be largely independent of eccentricity for $e \lesssim 0.2$ \cite{Huerta:2019oxn}, in the limit of a head-on collision \cite{Anninos:1993zj} the resulting spin is of course zero. The largest dimensionless spins are seen in the third generation (mass $\sim 3m_0$) BHs, with $\chi \sim 0.5$ to $0.9$. This is again distinct from Newtonian simulations of much less compact clusters where in, e.g., \cite{Rodriguez:2019huv} the spins are $\chi \sim 0.39$ to $0.45$ for third generation BHs (inferred using models fit to numerical relativity). The overall dimensionless spin and mass evolutions for the main (runaway) merger chain are shown in Fig. \ref{fig:spin_vs_mass}. We see that the dimensionless spin of the largest BH generally decreases with increasing mass after the third generation, albeit with lots of scatter due to the random nature of the mergers. 

Gammie, Shapiro and McKinney (2003) \cite{Gammie:2003qi}, following Hughes and Blandford (2003) \cite{Hughes:2002ei}, modelled the spin evolution of a central, slowly-rotating ($\chi  \ll 1$) BH accreting many much smaller companions with isotropically distributed orbital angular momentum slowly inspiraling through a sequence of nearly circular orbits due to the loss of gravitational radiation. Provided $\chi \gg \sqrt{27/(7\sqrt{2})q} \sim 3.1 q^{1/2}$, where $q$ is the mass ratio of the smaller to larger BH, this predicts that the spin should decay with increasing BH mass as $\chi \sim m^{-7/3}$ (shown as a grey dashed line in Fig. \ref{fig:spin_vs_mass}). At late times, when $\chi \lesssim 3.1 q^{1/2}$, they predict that $\chi$ will fluctuate around $q^{1/2}$. In our simulation we do indeed see that multiple mergers cause $\chi$ to decrease, confirming a prediction of \cite{Gammie:2003qi}. However, from Fig. \ref{fig:spin_vs_mass} one can see the decrease in spin from third generation onwards more closely approximates $\chi \sim q^{1/2} \sim (m/m_0)^{-1/2}$ (grey dotted line) rather than the steeper $\chi \sim m^{-7/3}$ decay (grey dashed line). This should not be surprising, given that for the merger with the third generation BH we already have $3.1 q^{1/2} \sim 3.1(1/3)^{1/2} \sim 1.8$ so $\chi < 3.1 q^{1/2}$. In addition the mergers in our simulations are highly eccentric, or direct collisions, rather than slow quasicircular inspirals, the third generation BHs are highly spinning, and we only have 25 objects at most.    

\subsection{What is the final fate of a bound, relativistic BH cluster?}

\begin{figure}
    \centering
    \includegraphics[width=\linewidth]{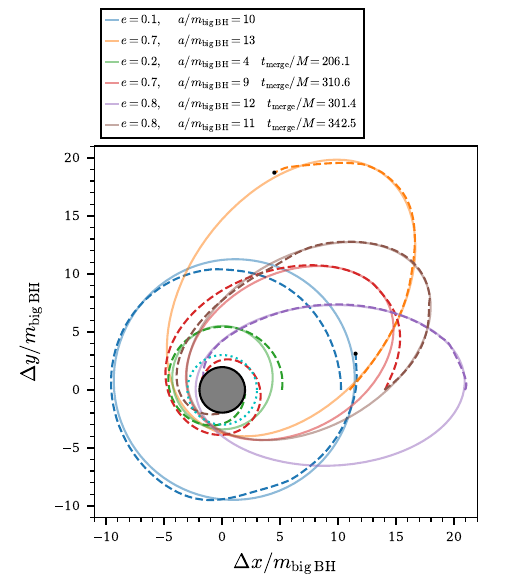}
    \label{ellipse}
    \caption{Late-time trajectories of the two remaining bound small BHs, along with the last four small BHs to merge, from $t/M = 151$ to $t/M = 347$. All trajectories have been rotated and projected such that they all lie in the same $xy$ plane. Treating the moving puncture coordinates as isotropic, and neglecting the small spin and velocity of the central BH, we also convert to (approximate) Schwarzschild coordinates by rescaling the radial distances from the central BH as $r_s = r\left(1 + \frac{m_{\mathrm{big\;BH}}}{2r}\right)^2$, where $r$ is the simulation ($\sim$isotropic) radius, $m_{\mathrm{big\;BH}}$ the final mass of the central BH and $r_s$ the areal radius. The numerical data is shown with black dashed lines, and the solid colored lines are Keplerian $r_s = a(1-e^2)/(1+e\cos(\phi-\phi_0))$ ellipses fit to the data. The grey circle shows the central BH, and the cyan dotted circle shows the photon sphere (again neglecting the BH spin).}
    \label{fig:final orbits}
\end{figure}

The state of the simulation at $t/M = 347$ (the end point of the simulation) is one large, central BH of mass $m \sim 22 m_0$, dimensionless spin $\chi \sim 0.3$ and apparent horizon velocity $v_{\mathrm{AH}} \sim 0.05$, with two first generation (mass $m_0$) BHs in bound orbits around it, and one first generation BH escaping to infinity. Fitting the orbits to the closest Keplerian ellipse, we find the small BHs in bound orbits have eccentricities $\sim 0.1$ and $0.7$ with semi-major axes $\sim 11$ and $15$ times the mass of the central, large BH. In Fig.\ref{fig:final orbits} we show these orbits (rotated and projected onto the same 2D plane), along with the orbits of the last four BHs to merge, from $t/M \sim 151$ to $t/M \sim 347$, and the corresponding best-fit Keplerian orbits. Making several further simplifying assumptions (using the quadrupole formula to estimate the gravitational luminosity and merger time, the Keplerian formula for the orbital period, neglecting the spin of the large BH, ignoring perturbations from the other small BHs and the small change in the large BH's mass by one or two additional mergers during the late-time orbit), we can estimate that the remaining bound BH with eccentricity $\sim 0.7$ will merge in $\lesssim 1$ orbit ($\lesssim 200M$) like the other mergers shown in Fig. \ref{fig:final orbits}, while the the BH with eccentricity $\sim 0.1$ should undergo more than 20 orbits (for a time of $\sim 4000M$) before merging. These late time mergers are more widely separated in time and therefore produce more easily distinguished inspiral, merger and ringdown binary merger gravitational wave signals which may be detected via matched-filtering searches, although we note that the generically large eccentricities will likely require eccentric waveform templates to avoid systematic biases in the parameter inference \cite{Favata:2021vhw,Saini:2022igm,Bhat:2024hyb}.  

With a mass ratio of around 1 to 20 these mergers may be classed as ``Intermediate Mass Ratio Inspirals" (IMRIs) (see e.g. \cite{Mandel:2007hi}). For a more massive cluster we can anticipate that the late stages of hierarchical growth will similarly produce ``Extreme Mass Ratio Inspirals" or EMRIs \cite{Amaro-Seoane:2007osp}, a key target for LISA \cite{Babak:2017tow}, which may offer a window into testing the astrophysical environment around the central BH \cite{Speri:2022upm,Dyson:2025dlj} and thus of the prior cluster.  

\section{Conclusions}

The dynamics of relativistic, strongly collisional, $N$-body systems of compact objects is a rich yet underexplored area of research. Extensive Newtonian and post-Newtonian simulations (e.g. \cite{Rodriguez:2017pec,Rodriguez:2019huv,Rizzuto:2021,Breen:2013vla,Barber:2024,Sedda:2024,Gaete:2024ovu}) have offered many valuable insights into IMBH formation and hierarchical mergers in dense star clusters, yet they have limited ability to describe highly relativistic systems as well as the gravitational waves emitted in clusters. 

In this work we pushed the limits of numerical relativity to achieve the first, preliminary, simulation of a idealized compact ($R/M \sim 10$) $N$-body cluster of 25 equal-mass nonspinning BHs in full GR. In Sec. \ref{sec:OOM} we discussed order of magnitude estimates for the key timescales for binary formation and BH mergers in the system, and from this predicted that the dominant formation channel for the first mergers would be direct collisions (unlike less compact systems with larger $N$ where quasi-circular binary formation would dominate), that the escape velocity of the clusters is sufficient for daughter BHs to be retained in the cluster, and that hierarchical runaway growth is typical. Several additional $N$-body simulations we have performed with different particles sampled from the same collisionless model all reveal these general features.

We constructed a bound, compact cluster, stable against immediate gravitational collapse or dispersal, following the prescription of Shapiro \& Teukolsky for relativistic, stable, monoenergetic, collisionless clusters in the smooth $N \rightarrow \infty$ limit \cite{Shapiro:1985a,Shapiro:1985b,Shapiro:1985c,Shapiro:1986}. By sampling 25 BHs obeying this distribution and evolving this system for up to $t \sim 347M$, we were able to make a first attempt at answering the questions posed in Sec \ref{sec:dyn_context}. The first BH mergers did indeed occur via direct collisions as we expected, in a time roughly comparable to our $\sim 28 M$ prediction from the order of magnitude estimate. At first there are multiple BH mergers distributed through the cluster. However, once a fourth generation BH (with roughly four times the mass of the initial BHs) is formed it undergoes runaway growth with a rapid sequence of mergers until it dominates the cluster. This central BH has accumulated a total of 22 of the original 25 BHs by the end point of the simulation at $t \sim 347M$. The final state includes two remaining bound first generation BHs, which will eventually merge with the central BH, and one ejected via a 4-body interaction with a large asymptotic velocity of $v_{\infty} \sim 0.51c$. We confirmed, for the first time, the predictions of \cite{Hughes:2002ei,Gammie:2003qi} showing that the spin of the central BH decreases with increasing mass from the fourth generation onwards. However, we only witnessed the $\chi \sim m^{-1/2}$ regime described in \cite{Gammie:2003qi} rather than the steeper $\chi \sim m^{-7/3}$ decay due to the (relatively) large initial mass ratio at the start of runaway growth.     

Finally, we were able to extract and analyse the gravitational wave signal. This showed several distinct features associated with the cluster regime. In the first $\sim 150M$ of simulation time the mergers are predominantly highly eccentric direct collisions, so the largest amplitude angular mode is $l  = 2, m = 0$ rather than the $l = m = 2$. In addition, the gravitational energy spectrum shows the $\dd E_{\mathrm{GW}}/\dd\omega \propto \omega^{4/3}$ power law expected at low frequencies for the collapse of initially non-relativistic systems and highly eccentric 2-body captures \cite{Wagoner:1979,Shapiro:1983} in contrast to the $\omega^{-1/3}$ behaviour expected for quasicircular low eccentricity binary mergers. In the regime of rapid runaway growth from $t/M \sim 55~-~90$, mergers and close hyperbolic and parabolic encounters are sufficiently frequent that the waveforms from individual events overlap, producing the largest gravitational wave luminosity of up to $\gtrsim 10^{56}\mathrm{erg}\;\mathrm{s}^{-1}$. At later times the merger events are more widely spaced as the number of bound BHs is depleted, producing eccentric, large mass ratio ``IMRI" waveforms. Comparing the frequency-domain characteristic strain to that of a equal-mass, quasi-circular binary BH merger with the same initial BH mass we found that the average cluster signal shows a lower amplitude at high frequencies but a larger amplitude at lower frequencies, with a peak characteristic strain amplitude at $f \sim 0.12/M$. As $N$-body systems are highly chaotic, making it difficult to produce exact templates for all possible real signals, the most realistic detection strategy is likely a burst search for unmodelled or partially modelled signals (e.g. \cite{Powell:2024bnp,KAGRA:2021tnv,Knee:2024mst}). A detailed analysis with a real burst search pipeline is beyond the scope of this preliminary study. However, using the signal to noise (SNR) values that can be obtained from a matched-template filter as a rough guide to detectability we find that next generation gravitational wave observatories should be able to detect signals from compact BH clusters across most of the observable universe. Advanced LIGO \cite{aLIGO:2020wna}, the Einstein Telescope \cite{Punturo:2010zz} and Cosmic Explorer \cite{Reitze:2019iox} will be most sensitive to clusters of stellar-origin BHs of initial mass $m_0 \sim 10 M_{\odot}$, while LISA \cite{LISA:2017} will be able to explore hierarchical growth in compact clusters of intermediate and supermassive BHs mass $m_0 \sim 10^{4} - 10^{6}M_{\odot}$, with SNR values of up to $\sim 80$ even at redshift 15. If such signals can be identified they may offer unique insights into galaxy formation in the early universe, and the formation of the early SMBHs detected by the James Webb Space Telescope (JWST) \cite{Banados:2017unc,Eilers:2024xus,Greene:2024phl,Sobolenko:2021orc,Kritos:2024sgd}.  

This work is, of course, merely the first foray into field of compact object collisional $N$-body clusters in numerical relativity, and there remains a huge amount of parameter space to explore. In future works we plan to study the impact of adding initial spin to the BHs, examine different numbers of BHs $N$ and different values of cluster compactness $R/M$, and try including a more realistic initial BH mass distribution appropriate for the different astrophysical scenarios. We also seek to explore environmental effects of the cluster from dark matter or baryonic gas, as the dynamical friction  and accretion from such environments may be very important in driving BH subsystems to highly compact states and therefore driving hierarchical growth. 

Movies of our cluster simulation and the associated gravitational waveform can be viewed at \cite{website}.

\section{Acknowledgements}
It is a pleasure to thank Robert Wald for several enlightening exchanges, and Josu Aurrekoetxea, Katy Clough and Pablo Galaviz for useful technical advice. We also thank members of our Illinois Relativity Undergraduate Research Team (Seyed Ahmad Dastgheib, Yuheng Guo, Yinuan Liang, Rohan Narasimhan and Cody Olson) for assistance with the 3D visualizations. This work was supported in part by National Science Foundation (NSF) Grants PHY-2308242, OAC-2310548 and PHY-2006066 to the University of Illinois at Urbana-Champaign. 
M.R. acknowledges support by the Generalitat Valenciana Grant CIDEGENT/2021/046, by the Spanish Agencia Estatal de Investigaci\'on (Grant PID2021-125485NB-C21) 
and  by the European Horizon Europe staff exchange (SE) program HORIZON-MSCA2021-SE-01 Grant No. NewFunFiCO-101086251.
A.T. acknowledges support from the National Center for Supercomputing Applications (NCSA) at the University of Illinois at Urbana-Champaign through the NCSA Fellows program. This work used Frontera at the Texas Advanced Computing Center (TACC) through allocation AST20025, MareNostrum 5 at the Barcelona Supercomputing Center (AECT-2023-1-0006) and Anvil at Purdue University through allocation MCA99S008 from the Advanced Cyberinfrastructure Coordination Ecosystem: Services \& Support (ACCESS) program, which is supported by National Science Foundation grants \#2138259, \#2138286, \#2138307, \#2137603, and \#2138296. Frontera is funded by the NSF through award \#1818253.

\appendix 

\section{Constructing a relativistic equilibrium cluster of N BHs.}
\label{app:construction_method}

To construct our initial data for a relativistic stable equilibrium cluster with a distribution
of $N$ equal rest-mass particles we begin by following the approach of 
Shapiro and Teukolsky \cite{Shapiro:1985b} for a collisionless system. 
Specifically, we first construct a smooth,  spherical, collisionless model (which corresponds to the limit of $N \rightarrow \infty$) for a monoenergetic distribution function of particles.  We write the spherically symmetric line element in Schwarzschild coordinates as 
\begin{equation}
    \dd s^2 = - e^{2\Phi} \dd t^2 + e^{2\Lambda}\dd r^2 + r^2 \dd \Omega^2,
\end{equation}
where $\dd \Omega^2 = r^2(\dd \theta^2 + \sin^2 \theta \dd \phi^2)$. The equations for a spherical cluster in equilibrium are
\begin{align}
    e^{2\Lambda} =& \left(1 - 2M(r)/r\right)^{-1}, \\
    \dv{M(r)}{r} =& \;4 \pi r^2 \rho, \label{eq:dMdr} \\
    \dv{\Phi(r)}{r} =& \frac{M(r)+4\pi r^3 P}{r(r - 2M)}, \\
    \dv{P(r)}{r} =& -\frac{(\rho+P)(M+4\pi r^3 P)}{r(r - 2M)}, \\
    \dv{M_0(r)}{r} =& \;4 \pi r^2 \rho_0, \label{eq:dM0dr}
\end{align} 
where $\rho, \rho_0, P$ are the smooth (mean-field) mass-energy density, rest-mass density, and pressure, $M_0(r)$ is the rest mass and $M(r)$ the gravitational mass within radius $r$. We adopt a monoenergetic distribution
\begin{equation}
    f(E) = K \delta(E - E_0),
\end{equation}
where $E$ is the ``energy at infinity" of a particle (= $-p_t$) and is everywhere equal to a constant $E_0$, and 
$K$ is a normalization constant. Then the moment equations (see, e.g.,\cite{Misner:1973prb}, Box 25.8) can be integrated analytically to give 
\begin{align}
    \rho =& \;4\pi K e^{2(\Phi_s-\Phi)} e^{-\Phi} \left[e^{2(\Phi_s-\Phi)} - 1\right]^{\frac{1}{2}}, \label{eq:rho_def} \\
    \rho_0 =& \;4\pi K e^{2(\Phi_s-\Phi)}e^{-\Phi_s}\left[e^{2(\Phi_s-\Phi)} - 1\right]^{\frac{1}{2}}, \\
    P =& \;\frac{4\pi K}{3} e^{-\Phi}\left[e^{2(\Phi_s-\Phi)} - 1\right]^{\frac{3}{2}}, \label{eq:P_def}
\end{align}
where $\Phi_s$ is the value of $\Phi$ at the surface where $P = \rho = 0$ so $e^{\Phi_s} = E_0/m_0$ where $m_0 = M_0 / N$. 

To integrate the cluster equilibrium equations we convert to dimensionless variables as in \cite{Ipser:1968a}. Let $P_c, \rho_c$ be the pressure and density at the center of the cluster at $r = 0$, then define
\begin{align}
    y :=& \;e^{2(\Phi - \Phi_s)}, \quad
    L := \left(\frac{P_c}{4\pi \rho^2_c}\right)^{\frac{1}{2}}, \;\;
    x := \; r/L, \notag \\
    v :=& \;\frac{M(r) \rho_c}{L P_c}, \quad
    v_0 := \frac{M_0(r) \rho_c}{L P_c}, \quad
    \; \zeta := \;P_c/\rho_c, \\
    \bar{\rho} :=& \;\rho/\rho_c, \quad\;\;\quad 
    \bar{\rho}_0 := \;\rho_0/\rho_c, \quad\quad\,
    \bar{P} := \;P/P_c, \notag
\end{align}
where $P_c, \rho_c$ are the pressure and density at the center of the cluster. Then $\zeta = \tfrac{1}{3} (1 - y_c)$ and Eqs. \ref{eq:dMdr}-\ref{eq:dM0dr} and \ref{eq:rho_def}-\ref{eq:P_def} become
\begin{align}
    \bar{\rho} = \left(\tfrac{y}{y_c}\right)^{-\frac{3}{2}} \left[\tfrac{y^{-1}-1}{y_c^{-1}-1}\right]^{\frac{1}{2}}, &\;\;   \bar{\rho}_0 = y^{\frac{1}{2}}_c \left(\tfrac{y}{y_c}\right)^{-1} \left[\tfrac{y^{-1}-1}{y_c^{-1}-1}\right]^{\frac{1}{2}}, \notag \\
    \bar{P} =& \left(\tfrac{y}{y_c}\right)^{-\frac{1}{2}}\left[\tfrac{y^{-1}-1}{y_c^{-1}-1}\right]^{\frac{3}{2}}, \label{eq:bar_defns} \\
    \dv{y}{x} =& \;2 \zeta \frac{y}{x^2}\frac{\left(v + \zeta x^3 \bar{P}\right)}{\left(1 - 2\zeta \frac{v}{x}\right)}, \\
    \dv{v}{x} =& \;x^2 \bar{\rho}, \\
    \dv{v_0}{x} =& \;x^2 \bar{\rho}_0 \left(1 - 2\zeta \frac{v}{x}\right)^{-\frac{1}{2}}, \label{eq:ODE_eqs}
\end{align}
where $y_c$ is the value of $y$ at $r = 0$. These equations can be integrated numerically with boundary conditions $y = y_c, v = v_0 = 0$ and $x = 0$ at the center and $y = 1$ at the surface, where we define $x(y=1) := x_s, v(y=1) := v_s, v_0(y=1) := v_{0,s}, r(y=1)=R$. The solutions in terms of the dimensionless quantities are entirely determined by the value of a single parameter, e.g. $y_c$. Useful ratios of physical quantities can then be obtained as 
\begin{equation}
\begin{split}
    R/M =& \frac{x_s}{v_s \zeta}, \quad \langle \rho \rangle/\rho_c = 3\frac{v_s}{x^3_s}, \quad M/m_0 = \frac{N v_s}{v_{0,s}}\\
    E_b/M =& \frac{M_0 - M}{M} = 1 - \frac{v_s}{v_{0,s}},
\end{split}
\end{equation}
where $E_b$ is the binding energy \cite{Shapiro:1985b}. 

To obtain an approximate equilibrium cluster with only $N$ discrete members we sample initial positions and 3-momenta so that the cluster tends to the equilibrium, smooth mean-field solution in the limit of large $N$. For the positions we draw a random sample with probability distribution $\mathcal{P}$ where the probability of finding a particle between $r$ and $r + \dd r$ is 
\begin{equation}
    \mathcal{P}(r)\dd r = \frac{4 \pi \rho_0(r) r^2}{m_0} \dd r 
\end{equation}
or in terms of $x$ 
\begin{equation}
    \mathcal{P}(x) \propto \rho_0(x) x^2 \propto y^{-1}\left[y^{-1} - 1\right]^{\frac{1}{2}} x^2. 
\end{equation}
We use a rejection technique to find $N$ values $\left\{x_i\right\}$ distributed according to this probability distribution for $i = 1 \dots N$. These are converted to radius values $\left\{r_i\right\}$ using $r_i/M = 3 x_i/(v_s (1 - y_c))$. We obtain angular positions $\left\{\theta_i,\phi_i\right\}$ from an isotropic distribution (i.e. a uniform distribution in $\phi$ and $\cos\theta$). Then in order to obtain initial data using the Bowen-York conformally flat prescription for $N$ BHs we need to transform from Schwarzschild coordinates $r,\theta,\phi$ to isotropic coordinates $\bar{r},\theta,\phi$ using 
\begin{equation}
    e^{2\Lambda} \dd r^2 + r^2 \dd \Omega^2 = \;A(\bar{r})^2\left(\dd \bar{r}^2 + \bar{r}^2 \dd \Omega^2\right),
\end{equation}
where matching the angular part yields $A(\bar{r}) = r/\bar{r}$ and the radial part gives
\begin{equation}
    \dv{\bar{r}}{r} = \frac{\bar{r}}{r}\left(1 - \frac{2M(r)}{r}\right)^{-\frac{1}{2}}\quad \textup{for}\;\; r < R, 
\end{equation}
or in convenient dimensionless units
\begin{equation}
    \dv{\ln(\bar{r}/M)}{\ln x} = \left[1 - 2\zeta \frac{v}{x}\right]^{-\frac{1}{2}}, \quad \textup{for}\;\; x < x_s. 
\end{equation}
Integrating (A18) in the exterior yields
\begin{equation}
    \bar{R}/M = \frac{1}{2}\left(\frac{R}{M}-1 + \left[\frac{R}{M}\left(\frac{R}{M}-2\right)\right]^{\frac{1}{2}}\right).
\end{equation}
Integrating (A18) numerically to map $r_i$ into 
$\bar r_i$ we then convert to isotropic Cartesian coordinates $\bar{x},\bar{y},\bar{z}$ as \mbox{$\bar{x}_i$ = $\bar{r}_i \sin \theta_i \cos \phi_i$, $\bar{y}_i$ = $\bar{r}_i \sin \theta_i \sin \phi_i$, $\bar{z}_i$ = $\bar{r}_i \cos \theta_i$} and assign positions $\boldsymbol{\bar x_i}/M = (\bar{x}_i/M,\bar{y}_i/M,\bar{z}_i/M)$.

Now we need the 3-momenta. The magnitude of the three-velocity $\vert\hat{v}\vert$ as observed by a static observer in 
a local orthnonormal frame $x^{\hat{\mu}}$ is given via
$p^{\hat{t}} = m_0/\sqrt{1 - \hat{v}^2}$ which yields 
$\vert\hat{v}\vert^2 = 1 - y$. The magnitude of the 3-momenta in the local orthnonormal coordinates is then
\begin{equation}
    \vert\hat{p}\vert = m_0 \gamma \hat{v} = m_0 \left[y^{-1} - 1\right]^{\frac{1}{2}}.
\end{equation}
The momenta should be isotropically distributed, so we select a new set of angles $\hat{\theta}_i,\hat{\phi}_i$ randomly distributed on the sphere and set $p^{\hat{x}}_i = \vert\hat{p}\vert_i \sin \hat{\theta}_i \cos \hat{\phi}_i,\; p^{\hat{y}}_i = \vert\hat{p}\vert_i \sin \hat{\theta}_i \sin \hat{\phi}_i,\; p^{\hat{z}}_i = \vert\hat{p}\vert_i \cos \hat{\theta}_i$. Finally, we obtain the 3-momenta in isotropic coordinates $\bar{p}^{a}_i$ for $\bar{a} = \bar{x},\bar{y},\bar{z}$ using $ p_i^{\bar{a}} = p_{\hat{i}}^{a}/A(\bar{r}_i)$ and then set $\bar{\boldsymbol{p}}_i = (p_i^{\bar{x}},p_i^{\bar{y}},p_i^{\bar{z}})$. We now have a suitable set of BH positions and momenta to input into our initial data solver.

\section{Tagging criteria and grid setup}
\label{app:grids}

As discussed in section \ref{sec:sim_setup}, we want our numerical grids to satisfy several requirements to ensure sufficient resolution in regions of interest. In the moving puncture gauge the radius of the apparent horizon of a black hole is roughly equal to the mass. The spatial resolution on the $l^{\mathrm{th}}$ refinement level is $\dd x = 2^{-l} \dd x_0$ where $\dd x_0$ is the spatial resolution on the coarsest level. Hence, in order to achieve the first requirement stated in \ref{sec:sim_setup}, a minimum $\mathcal{N}_{\mathrm{min}}$ number of grid points across each black hole horizon diameter, we determine a minimum refinement level required for the horizon of each BH as 
\begin{equation}
    l_{\textup{BH}} =  \textup{ceil}\left(\frac{\ln(\mathcal{N}_{\textup{min}} 2 m_{\textup{BH}} / \dd x_0)}{\ln(2)}\right),
\end{equation}
where ``$\mathrm{ceil}(x)$" denotes the ceiling function which returns the smallest integer greater than $x$ and  $m_{\textup{BH}}$ is the mass of the BH (which may be bigger than $m_0 = M_0/N$ if the BH is the product of mergers). We use $\mathcal{N}_{\mathrm{min}} = 32$, which we have found from experience to be the minimum required to evolve black holes accurately in \textsc{GRChombo}. The diameter of the smallest black hole horizon in the simulation is $\sim 2m_0$ where $m_0$ is the initial black hole mass, hence the spatial resolution (the $\dd x$ on the finest AMR level) is $\dd x_{\mathrm{min}} = 2m_0/32 = M/384$ where $M$ is the total ADM mass, as stated in Sec. \ref{sec:sim_timesteps}.

To achieve the second requirement, sufficiently spaced out level boundaries, we then tag cells on levels $l < l_{\textup{BH}}$ (to be refined on level $l$ + 1) if
\begin{equation}
    d_{\textup{BH}} < (1 + b)m_{\textup{BH}}\;2^{\textup{min}(l_{\textup{BH}}-l-1,2)}, 
\end{equation}
where $b = 0.5$ is a buffer factor (compare with Eqn. (45) in \cite{Radia:2021smk}) and we use 
\begin{equation}
    d_{\textup{BH}} = \textup{max}(\vert x - x_{\textup{BH}} \vert,\vert y - y_{\textup{BH}} \vert,\vert z - z_{\textup{BH}} \vert),
\end{equation}
(see Eqn. (46) in \cite{Radia:2021smk}) to ensure that the grid boundaries near the horizons are rectangular in order to minimize the amount of unnecessary grid refinement and de-refinement (here $x_\textup{BH},y_\textup{BH},z_\textup{BH}$ are the Cartesian coordinates of the BH). For the $t=0$ grids, we set $m_{\textup{BH}} = m_0$ for all BHs. 

In order to achieve the third requirement, sufficient resolution covering the extraction surfaces, we impose the extraction tagging criterion Eqn. (55) in \cite{Radia:2021smk} with a spatial resolution of $\dd x_{\textup{ex}} = 8 m_0 \sim 0.3 M $ in the extraction region. We use a cubic simulation domain of side length $L = 8192 m_0 \sim 340 M$ and a spatial resolution on the coarsest level of $\dd x_0 = L/256 = 32 m_0 = 1.3 M$. This means we end up needing 11 levels of AMR refinement to resolve the BH horizons. 

\section{Numerical convergence}
\label{app:convergence}

\begin{figure}
    \centering
    \includegraphics[width=\linewidth]{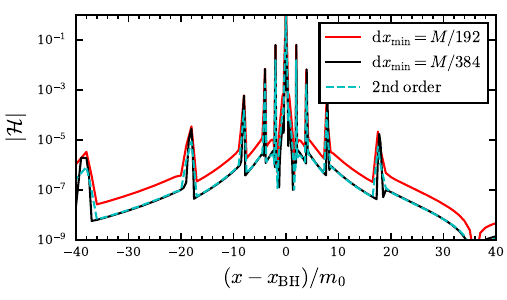}
    \caption{Convergence test for the initial data. We show the value of the Hamiltonian constraint error around one of the 25 BHs (representative of the other 24) on a log scale for a simulation with $dx_{\mathrm{min}} = M/192$ (red line) and one with twice the resolution $dx_{\mathrm{min}} = M/384$ (black line). The cyan dashed line shows the constraint error $\vert \mathcal{H} \vert^{\mathrm{ideal}}_{M/394}  = (192/384)^{2}\vert \mathcal{H} \vert_{M/192}$ one would expect for the higher resolution data if we had perfect 2nd order convergence (the order of the stencils in \textsc{GRTresna} \cite{Aurrekoetxea:2025kmm}).}
    \label{fig:initial_data_conv_test}
\end{figure}

\begin{figure}
    \centering
    \includegraphics[width=\linewidth]{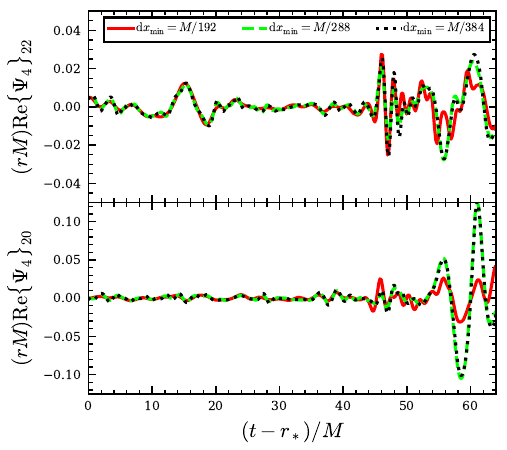}
    \caption{The real part of the $l=2,m=2$ and $l=2,m=0$ spheroidal harmonic components of $\Psi_4$, extracted on a spherical surface of $r_{\mathrm{ex}} = 25M$, for simulations using three different spatial resolutions. One can see that both sets of waveforms converge with increasing resolution, with the highest resolution of $\dd x_{\mathrm{min}} = M/384$ being the resolution used for the main simulation reported in this work.}
\label{fig:Psi_4_convergence}
\end{figure}

In Fig. \ref{fig:initial_data_conv_test} we show the magnitude of the Hamiltonian constraint for the initial data at different spatial resolutions on a line parallel to the $x$ axis passing through one of the BHs. The red and black lines show the value for the low and high resolutions respectively, and the cyan dashed line shows the value one would expect for the high resolution given the value for the low resolution and perfect $2^{\mathrm{nd}}$ order convergence (the order of the stencils used in \textsc{GRTresna}). As one can see the black and cyan dashed lines overlap, confirming we achieve the expected convergence order. In Fig. \ref{fig:Psi_4_convergence} we demonstrate the convergence of the $\Psi_4$ waveforms. While the lowest resolution of $\dd x_{\mathrm{min}} = M/192$ shows substantial deviations at later times, the two highest resolutions remain in close agreement, indicating that our $\dd x_{\mathrm{min}} = M/384$ simulation described in the main text is in the convergent regime and that the waveforms obtained are, in general, reliable. Computational resources limit the extent of the convergence studies we can perform in this preliminary study, however we plan to make a more thorough investigation of the effect of numerical resolution, gauge choices and different random draws for the initial BH positions and momenta in future works.

\section{Spatially-dependent gauge prescription}
\label{sec:eta_prescription}

We adopt the Gamma-driver shift condition
\begin{align}
    \partial_t \beta^i =& \;\tfrac{3}{4}B^i, \\
    \partial_t B^i =& \;\partial_t \hat{\Gamma}^i - \eta B^i,
\end{align}
as described in \cite{Radia:2021smk} Eqs. (30)-(31), but in order to better cope with the unequal-mass mergers that occur later in the cluster evolution we adopt a spatially-dependent function for $\eta$, following M\"{u}ller et al. (2010) \cite{Muller:2010zze} (Eqn. 5), with 
\begin{equation}
    \eta(\boldsymbol{r}) = \eta_0 + C\sum^N_{i=1} \frac{1/m_i - \eta_0}{1 + w_i \hat{r}^2_i}, 
\end{equation}
where $m_i$ is the mass of the $i^{\textup{th}}$ BH, $N$ is the number of BHs, and $\hat{r}_i = r_i/m_0$ where $r_i$ is the coordinate distance to the $i^{\textup{th}}$ BH and $m_0$ is the mass of the smallest BH (and the initial BH mass). We set $\eta_0 = 2/M$, where $M$ is the ADM mass of the whole system, and, for simplicity, set $C = 1$ and $w_i = w = 0.25$.  

\section{Escaping BH diagnostic}
\label{sec:escaping_BH_diagnostic}

Far outside the bound BH cluster we can approximate the spacetime in which an escaping BH of mass $m_0 \ll M$ moves by a Schwarzschild metric with mass $M_* = M - m_0$,
\begin{equation}
   \dd s^2 = -B \dd t^2 + B^{-1} \dd r^2 + r^2(\dd \theta^2 + \sin^2 \theta \dd \phi^2),  
\end{equation}
where $B = \left(1 - 2M_*/r\right)$. We can also approximate an escaping BH as a point particle moving along a geodesic, for which there is a constant of motion
\begin{align}
    -p_t = E = const,
\end{align}
where $p^{\mu}$ is its 4-momentum (see \cite{Shapiro:1983}, Eqn. 12.4.8) and $E$ its ``energy at infinity". Hence if $E/m_0 > 1$ the particle is unbound. The energy of the escaping BH as measured by a static local observer is then 
\begin{align}
\label{E}
    E_{\textup{local}} = \frac{E}{(1-2M_*/r)^{1/2}}
\end{align}
(see \cite{Shapiro:1983} Eqn. 12.4.9). We can equate $E_{\textup{local}}$ with $m_0\gamma = m_0 (1 - v^2)^{-\frac{1}{2}}$ where $v$ is the locally
measured 3-velocity. Taking $r \rightarrow \infty$ gives $E = m_0(1 - v_{\infty}^2)^{-\frac{1}{2}}$, hence substituting for $E$ and $E_{\textup{local}}$ in Eqn. \ref{E} gives
\begin{align}
\label{v}
   v^2(r) = v_{\infty}^2 + (1 - v_{\infty}^2)\frac{2M_*}{r},
\end{align}
obtaining the result in Eqn. \eqref{eq:Sch_v2}. To obtain $v_{\infty}$ we can rearrange Eqn. \ref{v} to give
\begin{equation}
    v_{\infty}^2 = \left(v^2 - \frac{2M_*}{r}\right)\left(1 - \frac{2M_*}{r}\right)^{-1}.
\end{equation}

Hence measuring the escaping BH's velocity $v(r_0)$ at a
single radius $r_0$ far from the cluster determines its velocity at infinity.
Note that at this level of approximation the difference between
areal (Schwarzschild) and isotropic radius at large $r \gg M_*$ is negligible.

\bibliography{apssamp}

\end{document}